\documentclass[a4paper,fleqn,usenatbib]{mnras}

\usepackage[english]{babel}
\usepackage{graphicx}
\usepackage{fleqn}
\usepackage{amssymb}
\usepackage{amsmath}
\usepackage{booktabs}
\usepackage{hyperref}
\usepackage{threeparttable}
\usepackage{comment}
\usepackage{txfonts} %older fonts, for compatibility with arXiv

\usepackage{color, colortbl}

\renewcommand{\S}{Section}
\newcommand{\F}{Fig.}

\newcommand{\ve}[1]{\boldsymbol{#1}}
\newcommand{\unit}[1]{\hat{\boldsymbol{#1}}}

\title[Chaos in quadruples]{Secular chaotic dynamics in hierarchical quadruple systems, with applications to hot Jupiters in stellar binaries and triples}

\author[Hamers \& Lai]{Adrian S. Hamers$^{1}$\thanks{E-mail: hamers@ias.edu} and Dong Lai$^{2,1,3}$ \\
$^{1}$Institute for Advanced Study, School of Natural Sciences, Einstein Drive, Princeton, NJ 08540, USA \\
$^{2}$Cornell Center for Astrophysics and Planetary Science, Department of Astronomy, Cornell University, Ithaca, NY 14853, USA \\
$^{3}$Institute for Theory and Computation, Harvard-Smithsonian Center for Astrophysics, Cambridge, MA 02138, USA}
\date{Accepted 2017 May 23. Received 2017 May 20; in original form 2017 April 26}

\pubyear{2017}

\begin{document}

\onecolumn

\label{firstpage}
\pagerange{\pageref{firstpage}--\pageref{lastpage}}
\maketitle

\begin{abstract} 
Hierarchical quadruple systems arise naturally in stellar binaries and triples that harbour planets. Examples are hot Jupiters (HJs) in stellar triple systems, and planetary companions to HJs in stellar binaries. The secular dynamical evolution of these systems is generally complex, with secular chaotic motion possible in certain parameter regimes. The latter can lead to extremely high eccentricities and, therefore, strong interactions such as efficient tidal evolution. These interactions are believed to play an important role in the formation of HJs through high-eccentricity migration. Nevertheless, a deeper understanding of the secular dynamics of these systems is still lacking. Here we study in detail the secular dynamics of a special case of hierarchical quadruple systems in either the `2+2' or `3+1' configurations. We show how the equations of motion can be cast in a form representing a perturbed hierarchical {\it three-body} system, in which the outer orbital angular momentum vector is precessing steadily around a fixed axis. In this case, we show that eccentricity excitation can be significantly enhanced when the precession period is comparable to the Lidov-Kozai oscillation time-scale of the inner orbit. This arises from an induced large mutual inclination between the inner and outer orbits driven by the precession of the outer orbit, even if the initial mutual inclination is small. We present a simplified semi-analytic model that describes the latter phenomenon.
\end{abstract}

\begin{keywords}
planets and satellites: dynamical evolution and stability -- planet-star interactions -- gravitation
\end{keywords}

\section{Introduction}
\label{sect:introduction}
\begin{figure*}
\centering
\includegraphics[scale = 0.55, trim = 0mm 20mm 0mm 0mm]{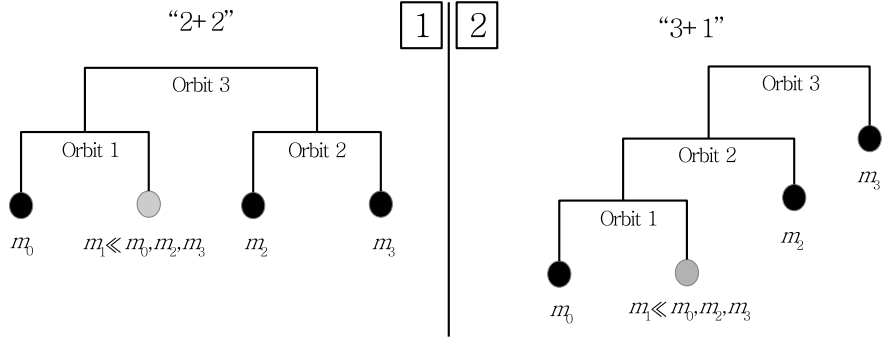}
\caption{\small Schematic representations of hierarchical orbits of test particles (bodies labelled $m_1$) in self-gravitating triple systems, using mobile diagrams \citep{1968QJRAS...9..388E}. Note that these diagrams only depict the hierarchy of the system, and not the relative sizes and orientations of the orbits. We consider two distinct configurations, shown in the first and second panels. With a number of approximations, the equations of motion for the two configurations are mathematically identical (see \S\,\ref{sect:model}).}
\label{fig:configurations}
\end{figure*}

Approximately 1 per cent of all stellar FG dwarf systems are hierarchical quadruples \citep{2014AJ....147...86T,2014AJ....147...87T}. In addition to these purely stellar systems, hierarchical quadruple configurations also occur naturally in lower multiplicity stellar systems that harbour planets. For example, there are currently three hot Jupiters (HJs; Jupiter-like planets orbiting stars in several days) known in stellar triple systems, i.e., WASP-12b \citep{2009ApJ...693.1920H,2013MNRAS.428..182B,2014ApJ...788....2B}, HAT-P-8b \citep{2009ApJ...704.1107L,2013MNRAS.428..182B,2014ApJ...788....2B} and KELT-4Ab \citep{2016AJ....151...45E}. In these systems, the HJ and its host star are orbited by a stellar binary (see the left-hand panel of \F\,\ref{fig:configurations}). Such a binary may have played a role in the formation and migration of the proto-HJ.

In particular, as shown by \citet{2017MNRAS.466.4107H}, the `binarity' of the companion can introduce secular enhancement of the eccentricity of the proto-HJ orbit in a larger parameter space compared to the situation when the star+HJ system is orbited by a single star. In the latter case, the eccentricity excitation is driven by Lidov-Kozai (LK) oscillations \citep{1962P&SS....9..719L,1962AJ.....67..591K} that arise in hierarchical three-body systems. Such an enhancement of the eccentricity excitation of proto-HJs in stellar triples compared to stellar binaries is relevant, because high-eccentricity migration models of HJs in stellar binaries \citep{2003ApJ...589..605W,2007ApJ...669.1298F,2012ApJ...754L..36N,2015ApJ...799...27P,2016MNRAS.456.3671A,2016MNRAS.460.1086M} are faced with the problem that the predicted formation rates are about 5-10 times lower than observed. One of the reasons for the lower rates is that short-range force precession in the orbit of the proto-HJ suppresses secular excitation if the orbit of the stellar companion is relatively wide \citep{2016ApJ...827....8N}.

Related to the above, the efficiency of high-eccentricity migration in stellar binaries could be enhanced if there are (currently undetected) massive planetary companions to HJs in stellar binaries, orbiting in-between the HJ and the stellar binary companion (see the right-hand panel of \F\,\ref{fig:configurations}). If the planetary companion satisfies several constraints, it can mediate LK-like oscillations in the proto-HJ orbit induced by the stellar companion, even if the planetary companion was initially coplanar with respect to the proto-HJ. This would put further constraints on high-eccentricity migration if, in the future, such currently unseen companions to HJs in stellar binaries are found to be absent \citep{2017ApJ...835L..24H}. 

The two configurations of HJs discussed above can be classified as `2+2' and `3+1' quadruple systems (see \F\,\ref{fig:configurations}). Although the orbit-averaged Hamiltonian and the equations of motion for these systems (and higher multiplicity systems) are known (\citealt{2015MNRAS.449.4221H,2016MNRAS.461.3964V,2016MNRAS.459.2827H}; see also, e.g., \citealt{2015ApJ...805...75P,2015MNRAS.447..747L} for vector-form equations for triples), a deeper understanding of the underlying mechanism for the enhanced eccentricity excitations, and the associated chaotic dynamics, is currently lacking. 

In this paper, we study in detail the secular dynamics of low-mass objects (planets or HJs) in quadruple systems with three more massive bodies. We show that, with a number of additional assumptions, the equations of motion for the `2+2' and `3+1' configurations can be cast into a single, general, form that is mathematically identical to that of a hierarchical three-body system in which the outer orbital angular momentum vector is precessing steadily around a fixed axis (\S\,\ref{sect:model}). Although the latter model is not amenable to analytic solutions, we show qualitatively how eccentricity excitation arises in this model (\S\,\ref{sect:num_gen}). In addition, we present another simplified model for which analytic results can be obtained, and which provides physical insight (\S\,\ref{sect:incl_model}). We conclude in \S\,\ref{sect:conclusions}.

Nearing the completion of this paper, we became aware of the simultaneous work of \citet{2017arXiv170505848P}, who discuss similar dynamics of perturbed hierarchical three-body systems in a different context. \citet{2017arXiv170505848P} consider binaries embedded in a non-spherical nuclear star cluster, and find that extreme eccentricity excitation is possible if the LK time-scale associated with the torque of the central massive black hole is comparable to the nodal precession time-scale of the binary centre of mass associated with the nuclear star cluster.  

\section{Model}
\label{sect:model}
In this section, we present a simplified model for the secular dynamics of hierarchical quadruple systems in which a test particle is orbiting in a system with three more massive bodies. In hierarchical triple systems, a useful approximation is the quadrupole-order test particle limit in which the angular momentum of the inner orbit, $L_\mathrm{in}$, is negligible compared to the angular momentum of the outer orbit, $L_\mathrm{out}$. In this limit, the outer orbital angular momentum vector $\ve{L}_\mathrm{out}$ is constant (therefore, $e_\mathrm{out}$ is constant as well), and the system is completely integrable. This system gives rise to well-known LK oscillations, occurring if $\ve{L}_\mathrm{in}$ and $\ve{L}_\mathrm{out}$ are initially inclined by more than $i_\mathrm{crit} \equiv \mathrm{arccos}(\sqrt{3/5}) \approx 39.2315^\circ$, and with a maximum eccentricity of $e_\mathrm{max} = [1- (5/3) (\unit{L}_\mathrm{in,init} \cdot \unit{L}_\mathrm{out})^2]^{1/2}$, where $\unit{L}_\mathrm{in,init}$ is the initial unit inner orbital angular momentum vector (assuming a zero initial eccentricity).

Our model applies to hierarchical {\it quadruple} systems, but we will show that it can be considered as a perturbed hierarchical {\it three-body} problem in the test particle limit. In our case, $\ve{L}_\mathrm{out}$ (of the perturbed three-body problem) is no longer constant but precesses steadily around a fixed axis. With the introduction of this perturbation, the system is no longer integrable, and the evolution is generally more complex. In particular, chaotic secular behaviour can be induced. Below, we will show that this perturbed model applies to two types of restricted hierarchical quadruple systems (the `2+2' and `3+1' configurations). Both cases have direct applications to planetary systems.

\subsection{2+2 quadruple systems}
\label{sect:model:2p2}
Consider hierarchical quadruple systems in the `2+2' configuration, i.e., two binaries orbiting each other's barycenter. The hierarchy and notation are indicated schematically in the left-hand panel of \F\,\ref{fig:configurations}. To the quadruple-order, i.e., the second order in the ratios of the orbital separations, the orbit-averaged Hamiltonian was derived by \citet{2015MNRAS.449.4221H} and \cite{2016MNRAS.459.2827H}, and consists of the hierarchical three-body Hamiltonian applied to the (1,3) orbit pair, plus the hierarchical three-body Hamiltonian applied to the (2,3) orbit pair. The equations of motion for the eccentricity $\ve{e}_i$ and angular-momentum $\ve{j}_i = \sqrt{1-e_i^2} \, \unit{L}_i$ vectors for the three orbits follow from the Milankovitch equations (\citealt{milankovitch_39,1961JGR....66.2797M,1963PCPS...59..669A,1964RSPSA.280...97A,2005MNRAS.364.1222B,2009AJ....137.3706T}), and read
\begin{subequations}
\label{eq:EOM_2p2_gen}
\begin{align}
\label{eq:EOM_2p2_gen:j1}
\frac{\mathrm{d} \ve{j}_1}{\mathrm{d} t} &= \frac{3}{4} t_\mathrm{LK,2+2,13}^{-1} \left [ \left ( \ve{j}_1 \cdot \unit{L}_3 \right ) \left ( \ve{j}_1 \times \unit{L}_3 \right ) - 5 \left ( \ve{e}_1 \cdot \unit{L}_3 \right ) \left ( \ve{e}_1 \times \unit{L}_3 \right )  \right ]; \\
\label{eq:EOM_2p2_gen:e1}
\frac{\mathrm{d} \ve{e}_1}{\mathrm{d} t} &= \frac{3}{4} t_\mathrm{LK,2+2,13}^{-1} \left [ \left ( \ve{j}_1 \cdot \unit{L}_3 \right ) \left ( \ve{e}_1 \times \unit{L}_3 \right ) + 2 \left ( \ve{j}_1 \times \ve{e}_1 \right ) - 5 \left ( \ve{e}_1 \cdot \unit{L}_3 \right ) \left ( \ve{j}_1 \times \unit{L}_3 \right )  \right ]; \\
\frac{\mathrm{d} \ve{j}_2}{\mathrm{d} t} &= \frac{3}{4} t_\mathrm{LK,2+2,23}^{-1} \left [ \left ( \ve{j}_2 \cdot \unit{L}_3 \right ) \left ( \ve{j}_2 \times \unit{L}_3 \right ) - 5 \left ( \ve{e}_2 \cdot \unit{L}_3 \right ) \left ( \ve{e}_2 \times \unit{L}_3 \right )  \right ]; \\
\frac{\mathrm{d} \ve{e}_2}{\mathrm{d} t} &= \frac{3}{4} t_\mathrm{LK,2+2,23}^{-1} \left [ \left ( \ve{j}_2 \cdot \unit{L}_3 \right ) \left ( \ve{e}_2 \times \unit{L}_3 \right ) + 2 \left ( \ve{j}_2 \times \ve{e}_2 \right ) - 5 \left ( \ve{e}_2 \cdot \unit{L}_3 \right ) \left ( \ve{j}_2 \times \unit{L}_3 \right )  \right ]; \\
\label{eq:EOM_2p2_gen:j3}
\frac{\mathrm{d} \ve{j}_3}{\mathrm{d} t} &= \frac{3}{4} t_\mathrm{LK,2+2,13}^{-1} \frac{\Lambda_1}{\Lambda_3} \left [ 5 \left ( \ve{e}_1 \cdot \unit{L}_3 \right ) \left ( \ve{e}_1 \times \unit{L}_3 \right ) - \left ( \ve{j}_1 \cdot \unit{L}_3 \right ) \left ( \ve{j}_1 \times \unit{L}_3 \right )  \right ] + \frac{3}{4} t_\mathrm{LK,2+2,23}^{-1} \frac{\Lambda_2}{\Lambda_3} \left [ 5 \left ( \ve{e}_2 \cdot \unit{L}_3 \right ) \left ( \ve{e}_2 \times \unit{L}_3 \right ) - \left ( \ve{j}_2 \cdot \unit{L}_3 \right ) \left ( \ve{j}_2 \times \unit{L}_3 \right )  \right ]; \\
\label{eq:EOM_2p2_gen:e3}
\nonumber \frac{\mathrm{d} \ve{e}_3}{\mathrm{d} t} &= \frac{3}{8} t_\mathrm{LK,2+2,13}^{-1} \frac{\Lambda_1}{\Lambda_3} \frac{1}{\sqrt{1-e_3^2}} \left [  \left \{ \left (1-6e_1^2 \right ) + 25 \left ( \ve{e}_1 \cdot \unit{L}_3\right)^2 - 5 \left ( \ve{j}_1 \cdot \unit{L}_3 \right )^2 \right \} \left ( \ve{e}_3 \times \unit{L}_3 \right ) - 10 \left ( \ve{e}_1 \cdot \unit{L}_3 \right ) \left ( \ve{e}_3 \times \ve{e}_1 \right ) + 2 \left ( \ve{j}_1 \cdot \unit{L}_3 \right ) \left ( \ve{e}_3 \times \ve{j}_1 \right ) \right ]  \\
\nonumber &+ \frac{3}{8} t_\mathrm{LK,2+2,23}^{-1} \frac{\Lambda_2}{\Lambda_3} \frac{1}{\sqrt{1-e_3^2}} \left [  \left \{ \left (1-6e_2^2 \right ) + 25 \left ( \ve{e}_2 \cdot \unit{L}_3\right)^2 - 5 \left ( \ve{j}_2 \cdot \unit{L}_3 \right )^2 \right \} \left ( \ve{e}_3 \times \unit{L}_3 \right ) - 10 \left ( \ve{e}_2 \cdot \unit{L}_3 \right ) \left ( \ve{e}_3 \times \ve{e}_2 \right ) + 2 \left ( \ve{j}_2 \cdot \unit{L}_3 \right ) \left ( \ve{e}_3 \times \ve{j}_2 \right ) \right ] \\
&\equiv  \ve{e}_3 \times \ve{f}(\ve{e}_1,\ve{j}_1,\ve{e}_2,\ve{j}_2,\ve{j}_3).
\end{align}
\end{subequations}
Here, $\Lambda_i$ is the angular momentum of orbit $i$ for a circular orbit which is constant in the secular approximation, i.e., $\Lambda_i = \mu_i \sqrt{GM_i a_i}$, where $\mu_i$ and $M_i$ are the reduced and the total mass, respectively, of binary $i$. The (two) LK time-scales are given by
\begin{subequations}
\label{eq:2p2_t_LK}
\begin{align}
t_\mathrm{LK,2+2,13} &= \frac{m_0+m_1}{m_2+m_3} \sqrt{ \frac{a_1^3}{G(m_0+m_1)}} \left ( \frac{a_3}{a_1} \right )^3 \left ( 1 - e_3^2 \right )^{3/2}; \qquad t_\mathrm{LK,2+2,23} = \frac{m_2+m_3}{m_0+m_1} \sqrt{ \frac{a_2^3}{G(m_2+m_3)}} \left ( \frac{a_3}{a_2} \right )^3 \left ( 1 - e_3^2 \right )^{3/2}.
\end{align}
\end{subequations}
The function $\ve{f}$ in equation~(\ref{eq:EOM_2p2_gen:e3}) is independent of $\ve{e}_3$, showing that $\ve{e}_3$ precesses around $\unit{f}$ and the magnitude of $e_3$ remains constant. Therefore, the LK time-scales in equations~(\ref{eq:2p2_t_LK}) are constant as well.

The coupled equations~(\ref{eq:EOM_2p2_gen}) are generally not amenable to analytic solutions. We simplify them by assuming the test particle limit, $\Lambda_1 \ll \Lambda_2, \Lambda_3$, and setting $\ve{e}_2 = \ve{0}$. Evidently, the torque of orbit 3 can excite $e_2$ if orbits 2 and 3 are initially sufficiently inclined. Here, we assume that $i_\mathrm{23,init} \lesssim 40^\circ$, such that the LK mechanism is not active for the orbit pair (2,3); therefore, $e_2$ is constant. The assumption on the $\Lambda_i$ implies that the first term in equation~(\ref{eq:EOM_2p2_gen:j3}), proportional to $\Lambda_1/\Lambda_3$, is negligible compared to the second term, which is proportional to $\Lambda_2/\Lambda_3$ (also assuming that $t_\mathrm{LK,2+2,13}$ and $t_\mathrm{LK,2+2,23}$ are not too distinct). Equation~(\ref{eq:EOM_2p2_gen:j3}) then reads
\begin{align}
\label{eq:EOM_2p2_j3_stage1}
\frac{\mathrm{d} \ve{j}_3}{\mathrm{d} t} &\approx - \frac{3}{4} t_\mathrm{LK,2+2,23}^{-1} \frac{\Lambda_2}{\Lambda_3} \left ( \unit{L}_2 \cdot \unit{L}_3 \right ) \left ( \unit{L}_2 \times \unit{L}_3 \right ).
\end{align}
Using that the total angular momentum vector, $\ve{L}_\mathrm{tot} = \ve{L}_1 + \ve{L}_2 + \ve{L}_3$, is conserved, equation~(\ref{eq:EOM_2p2_j3_stage1}) can be written in the form
\begin{align}
\label{eq:EOM_2p2_j3_stage2}
\frac{\mathrm{d} \unit{L}_3}{\mathrm{d} t} &\approx - \ve{\Omega}_3 \times \unit{L}_3 ,
\end{align}
where $\ve{\Omega}_3$ is a constant vector with magnitude
\begin{align}
\label{eq:2p2_Omega_3}
\Omega_3 &=  \frac{3}{4} t_\mathrm{LK,2+2,23}^{-1}  \cos(i_{23,\mathrm{init}}),
\end{align}
with $i_{23,\mathrm{init}}$ as the initial inclination between $\unit{L}_2$ and $\unit{L}_3$.

In summary, the secular dynamics of the restricted problem, $\Lambda_1 \ll \Lambda_2, \Lambda_3$ and $\ve{e}_2 = \ve{0}$, are described by equations~(\ref{eq:EOM_2p2_gen:j1}), (\ref{eq:EOM_2p2_gen:e1}) and (\ref{eq:EOM_2p2_j3_stage2}). These restricted equations apply, e.g., to a planet orbiting a star (orbit 1) that is orbited by a more distant stellar binary (orbit 2), in a relatively wide orbit (orbit 3).

\subsection{3+1 quadruple systems}
\label{sect:model:3p1}
Next, we consider hierarchical quadruple systems in the `3+1' configuration, i.e., a triple orbited by a distant fourth body (see the second panel in \F\,\ref{fig:configurations}). To quadrupole order, the orbit-averaged Hamiltonian is given by adding the relevant Hamiltonians from the hierarchical three-body Hamiltonian, i.e., the three Hamiltonians associated with the (1,2), (2,3) and (1,3) pairs \citep{2015MNRAS.449.4221H,2016MNRAS.459.2827H}. The equations of motion read
\begin{subequations}
\label{eq:EOM_3p1_gen}
\begin{align}
\label{eq:EOM_3p1_gen:j1}
\frac{\mathrm{d} \ve{j}_1}{\mathrm{d} t} &= \frac{3}{4} t_\mathrm{LK,3+1,12}^{-1} \left [ \left ( \ve{j}_1 \cdot \unit{L}_2 \right ) \left ( \ve{j}_1 \times \unit{L}_2 \right ) - 5 \left ( \ve{e}_1 \cdot \unit{L}_2 \right ) \left ( \ve{e}_1 \times \unit{L}_2 \right )  \right ] + \frac{3}{4} t_\mathrm{LK,3+1,13}^{-1} \left [ \left ( \ve{j}_1 \cdot \unit{L}_3 \right ) \left ( \ve{j}_1 \times \unit{L}_3 \right ) - 5 \left ( \ve{e}_1 \cdot \unit{L}_3 \right ) \left ( \ve{e}_1 \times \unit{L}_3 \right )  \right ]; \\
\nonumber \frac{\mathrm{d} \ve{e}_1}{\mathrm{d} t} &= \frac{3}{4} t_\mathrm{LK,3+1,12}^{-1} \left [ \left ( \ve{j}_1 \cdot \unit{L}_2 \right ) \left ( \ve{e}_1 \times \unit{L}_2 \right ) + 2 \left ( \ve{j}_1 \times \ve{e}_1 \right ) - 5 \left ( \ve{e}_1 \cdot \unit{L}_2 \right ) \left ( \ve{j}_1 \times \unit{L}_2 \right )  \right ] + \frac{3}{4} t_\mathrm{LK,3+1,13}^{-1} \left [ \left ( \ve{j}_1 \cdot \unit{L}_3 \right ) \left ( \ve{e}_1 \times \unit{L}_3 \right ) + 2 \left ( \ve{j}_1 \times \ve{e}_1 \right ) \right. \\
\label{eq:EOM_3p1_gen:e1}
&\quad \left. - 5 \left ( \ve{e}_1 \cdot \unit{L}_3 \right ) \left ( \ve{j}_1 \times \unit{L}_3 \right )  \right ]; \\
\label{eq:EOM_3p1_gen:j2}
\frac{\mathrm{d} \ve{j}_2}{\mathrm{d} t} &= \frac{3}{4} t_\mathrm{LK,3+1,12}^{-1} \frac{\Lambda_1}{\Lambda_2} \left [ 5 \left ( \ve{e}_1 \cdot \unit{L}_2 \right ) \left ( \ve{e}_1 \times \unit{L}_2 \right ) - \left ( \ve{j}_1 \cdot \unit{L}_2 \right ) \left ( \ve{j}_1 \times \unit{L}_2 \right )  \right ] - \frac{3}{4} t_\mathrm{LK,3+1,23}^{-1} \left [ 5 \left ( \ve{e}_2 \cdot \unit{L}_3 \right ) \left ( \ve{e}_2 \times \unit{L}_3 \right ) - \left ( \ve{j}_2 \cdot \unit{L}_3 \right ) \left ( \ve{j}_2 \times \unit{L}_3 \right )  \right ]; \\
\nonumber \frac{\mathrm{d} \ve{e}_2}{\mathrm{d} t} &= \frac{3}{8} t_\mathrm{LK,3+1,12}^{-1} \frac{\Lambda_1}{\Lambda_2} \frac{1}{\sqrt{1-e_2^2}} \left [  \left \{ \left (1-6e_1^2 \right ) + 25 \left ( \ve{e}_1 \cdot \unit{L}_2\right)^2 - 5 \left ( \ve{j}_1 \cdot \unit{L}_2 \right )^2 \right \} \left ( \ve{e}_2 \times \unit{L}_2 \right ) - 10 \left ( \ve{e}_1 \cdot \unit{L}_2 \right ) \left ( \ve{e}_2 \times \ve{e}_1 \right ) + 2 \left ( \ve{j}_1 \cdot \unit{L}_2 \right ) \left ( \ve{e}_2 \times \ve{j}_1 \right ) \right ] \\
& + \frac{3}{4} t_\mathrm{LK,23}^{-1} \left [ \left ( \ve{j}_2 \cdot \unit{L}_3 \right ) \left ( \ve{e}_2 \times \unit{L}_3 \right ) + 2 \left ( \ve{j}_2 \times \ve{e}_2 \right ) - 5 \left ( \ve{e}_2 \cdot \unit{L}_3 \right ) \left ( \ve{j}_2 \times \unit{L}_3 \right )  \right ]; \\
\label{eq:EOM_3p1_gen:j3}
\frac{\mathrm{d} \ve{j}_3}{\mathrm{d} t} &= \frac{3}{4} t_\mathrm{LK,23}^{-1} \frac{\Lambda_2}{\Lambda_3} \left [ 5 \left ( \ve{e}_2 \cdot \unit{L}_3 \right ) \left ( \ve{e}_2 \times \unit{L}_3 \right ) - \left ( \ve{j}_2 \cdot \unit{L}_3 \right ) \left ( \ve{j}_2 \times \unit{L}_3 \right )  \right ] + \frac{3}{4} t_\mathrm{LK,13}^{-1} \frac{\Lambda_1}{\Lambda_3} \left [ 5 \left ( \ve{e}_1 \cdot \unit{L}_3 \right ) \left ( \ve{e}_1 \times \unit{L}_3 \right ) - \left ( \ve{j}_1 \cdot \unit{L}_3 \right ) \left ( \ve{j}_1 \times \unit{L}_3 \right )  \right ]; \\
\label{eq:EOM_3p1_gen:e3}
\frac{\mathrm{d} \ve{e}_3}{\mathrm{d} t} &= \frac{3}{8} t_\mathrm{LK,3+1,23}^{-1} \frac{\Lambda_2}{\Lambda_3} \frac{1}{\sqrt{1-e_3^2}} \left [  \left \{ \left (1-6e_2^2 \right ) + 25 \left ( \ve{e}_2 \cdot \unit{L}_3\right)^2 - 5 \left ( \ve{j}_2 \cdot \unit{L}_3 \right )^2 \right \} \left ( \ve{e}_3 \times \unit{L}_3 \right ) - 10 \left ( \ve{e}_2 \cdot \unit{L}_3 \right ) \left ( \ve{e}_3 \times \ve{e}_2 \right ) + 2 \left ( \ve{j}_2 \cdot \unit{L}_3 \right ) \left ( \ve{e}_3 \times \ve{j}_2 \right ) \right ] \\
& + \frac{3}{8} t_\mathrm{LK,3+1,13}^{-1} \frac{\Lambda_1}{\Lambda_3} \frac{1}{\sqrt{1-e_3^2}} \left [  \left \{ \left (1-6e_1^2 \right ) + 25 \left ( \ve{e}_1 \cdot \unit{L}_3\right)^2 - 5 \left ( \ve{j}_1 \cdot \unit{L}_3 \right )^2 \right \} \left ( \ve{e}_3 \times \unit{L}_3 \right ) - 10 \left ( \ve{e}_1 \cdot \unit{L}_3 \right ) \left ( \ve{e}_3 \times \ve{e}_1 \right ) + 2 \left ( \ve{j}_1 \cdot \unit{L}_3 \right ) \left ( \ve{e}_3 \times \ve{j}_1 \right ) \right ] \\
&\equiv \ve{e}_3 \times \ve{f}(\ve{e}_1,\ve{j}_1,\ve{e}_2,\ve{j}_2,\ve{j}_3).
\end{align}
\end{subequations}
The (three) LK time-scales are now given by
\begin{subequations}
\label{eq:3p1_t_LK}
\begin{align}
t_\mathrm{LK,3+1,12} &= \frac{m_0+m_1}{m_2} \sqrt{ \frac{a_1^3}{G(m_0+m_1)}} \left ( \frac{a_2}{a_1} \right )^3 \left ( 1 - e_2^2 \right )^{3/2}; \qquad t_\mathrm{LK,3+1,23}  = \frac{m_0+m_1+m_2}{m_3} \sqrt{ \frac{a_2^3}{G(m_0+m_1+m_2)}} \left ( \frac{a_3}{a_2} \right )^3 \left ( 1 - e_3^2 \right )^{3/2}; \\
t_\mathrm{LK,3+1,13} &= \frac{m_0+m_1}{m_3} \sqrt{ \frac{a_1^3}{G(m_0+m_1)}} \left ( \frac{a_3}{a_1} \right )^3 \left ( 1 - e_3^2 \right )^{3/2}.
\end{align}
\end{subequations}
Again, the function $\ve{f}$ in equation~(\ref{eq:EOM_3p1_gen:e3}) is independent of $\ve{e}_3$, showing that $\ve{e}_3$ precesses around $\unit{f}$ and $e_3$ is constant. In this case, the LK time-scale $t_\mathrm{LK,3+1,12}$ is generally not constant because $e_2$ is not generally constant.

We make the assumption $\Lambda_1 \ll \Lambda_2 \ll \Lambda_3$. Furthermore, we assume $\ve{e}_2 = \ve{0}$. As before, we assume that $i_\mathrm{23,init} \lesssim 40^\circ$ such that the LK mechanism does not operate for the orbit pair (2,3). Equation~(\ref{eq:EOM_3p1_gen:j2}) can then be written in the form
\begin{align}
\label{eq:3p1_dL2_dt}
\frac{\mathrm{d} \unit{L}_2}{\mathrm{d} t} &\approx -\ve{\Omega}_2 \times \unit{L}_2,
\end{align}
where $\unit{\Omega}_2 = \unit{L}_3$ is constant, and 
\begin{align}
\label{eq:3p1_Omega_2}
\Omega_2 &= \frac{3}{4} t_\mathrm{LK,3+1,23}^{-1}   \cos(i_{23,\mathrm{init}}).
\end{align}

In addition, for hierarchical systems ($a_3 \gg a_2$) with a sufficiently large ratio $m_2/m_3$, $t_\mathrm{LK,3+1,12} \ll t_\mathrm{LK,3+1,13}$, i.e., the torque of orbit 3 on orbit 1 is negligible compared to the torque of orbit 2 on orbit 1. The restricted problem is then defined by equations~(\ref{eq:EOM_3p1_gen:j1}) and (\ref{eq:EOM_3p1_gen:e1}), dropping terms proportional to $t_\mathrm{LK,3+1,13}^{-1}$, and equation (\ref{eq:3p1_dL2_dt}). It applies, e.g., to a massive planetary companion to a proto-HJ in a stellar binary.

\subsection{Generalized model}
\subsubsection{Equations of motion}
Although they correspond to different hierarchical configurations, the restricted equations of motion for the `2+2' and `3+1' configurations are mathematically identical. They can be interpreted as representing a perturbed hierarchical {\it three-body} system in the test particle limit, where the outer angular momentum vector is precessing around a fixed axis with a fixed rate $\Omega_\mathrm{out}$. The model can be written in the scaled form
\begin{subequations}
\label{eq:EOM_gen_scaled}
\begin{align}
\label{eq:EOM_gen_scaled:j}
\frac{\mathrm{d} \ve{j}}{\mathrm{d} \tau} &= \frac{3}{4}\left [ \left ( \ve{j} \cdot \unit{L}_\mathrm{out} \right ) \left ( \ve{j} \times \unit{L}_\mathrm{out} \right ) - 5 \left ( \ve{e} \cdot \unit{L}_\mathrm{out} \right ) \left ( \ve{e} \times \unit{L}_\mathrm{out} \right )  \right ]; \\
\label{eq:EOM_gen_scaled:e}
\frac{\mathrm{d} \ve{e}}{\mathrm{d} \tau} &= \frac{3}{4}\left [ \left ( \ve{j} \cdot \unit{L}_\mathrm{out} \right ) \left ( \ve{e} \times \unit{L}_\mathrm{out} \right ) + 2 \left ( \ve{j} \times \ve{e} \right ) - 5 \left ( \ve{e} \cdot \unit{L}_\mathrm{out} \right ) \left ( \ve{j} \times \unit{L}_\mathrm{out} \right )  \right ]; \\
\label{eq:EOM_gen_scaled:L_out}
\frac{\mathrm{d} \unit{L}_\mathrm{out}}{\mathrm{d} \tau} &= - \beta \left( \unit{z} \times \unit{L}_\mathrm{out} \right ).
\end{align}
\end{subequations}
Here, $\ve{e}$ and $\ve{j}$ correspond to the inner test-particle orbit, and $\tau \equiv t/t_\mathrm{LK}$ with the LK time-scale given by
\begin{align}
t_\mathrm{LK} = \frac{m_0+m_1}{m_\mathrm{out}} \sqrt{ \frac{a^3}{G(m_0+m_1)}} \left ( \frac{a_\mathrm{out}}{a} \right )^3 \left (1-e_\mathrm{out}^2 \right)^{3/2}.
\end{align}
The semimajor axis of the test-particle orbit is denoted with $a$, and the subscript `out' denotes properties of the `outer' companion, which depends on the hierarchical configuration (in the `2+2' case, $a_\mathrm{out}=a_3$, $e_\mathrm{out}=e_3$ and $m_\mathrm{out} = m_2+m_3$, whereas $a_\mathrm{out}=a_2$, $e_\mathrm{out}=e_2$ and $m_\mathrm{out} = m_2$ in the `3+1' case). The dimensionless and constant parameter $\beta$ is defined as
\begin{align}
\label{eq:beta_def}
\beta \equiv \Omega_\mathrm{out} \, t_\mathrm{LK}.
\end{align}
In equation~(\ref{eq:EOM_gen_scaled:L_out}), the fixed axis around which $\unit{L}_\mathrm{out}$ precesses is taken to be the $\unit{z}$ axis. We denote the angle between $\unit{L}_\mathrm{out}$ and $\unit{z}$ with $\alpha$.

\subsubsection{Hamiltonian and the `LK constant'}
If $\beta=0$ in equations~(\ref{eq:EOM_gen_scaled}), then the (conserved) Hamiltonian is mathematically equivalent to the three-body test-particle Hamiltonian. For nonzero $\beta$, this Hamiltonian is not constant because of the fixed precession of $\unit{L}_\mathrm{out}$. However, one can transform to a frame that is rotating around the $\unit{z}$ axis with the same rate at which $\unit{L}_\mathrm{out}$ is precessing around the $\unit{z}$ axis. This gives (e.g., \citealt{1993CeMDA..57..359K})
\begin{align}
\label{eq:H_gen_rot}
H_\mathrm{gen,rot} &= H_0 \left [ \left (1-6e^2 \right ) + 15 \left ( \ve{e} \cdot \unit{L}_\mathrm{out}\right)^2 - 3 \left ( \ve{j} \cdot \unit{L}_\mathrm{out} \right )^2 + 8 \beta  \left ( \ve{j} \cdot \unit{z} \right ) \right ],
\end{align}
where $H_0$ is a constant. 

In the test-particle three-body limit ($\beta=0$), the quantity $\sqrt{1-e^2} \cos(i)$, where $i$ is the mutual inclination between the inner and outer orbits, is a constant of the motion (known as the `LK' or `Kozai' constant). For $\beta \neq 0$, the `LK constant' is no longer conserved, but evolves according to
\begin{align}
\label{eq:LK_constant_nonzero}
\frac{\mathrm{d}}{\mathrm{d} t} \left [ \sqrt{1-e^2} \cos(i) \right ] = - \Omega_\mathrm{out} \, \ve{j} \cdot \left ( \unit{z} \times \unit{L}_\mathrm{out} \right ).
\end{align}

\section{Eccentricity excitation in the generalized model}
\label{sect:num_gen}
The secular three-body equations of motion in the quadrupole-order test particle limit are integrable, i.e., analytic solutions exist for the eccentricity as a function of time \citep{2007CeMDA..98...67K}. However, they are no longer amenable to simple solutions if the outer angular momentum vector is precessing as in our model described by equations~(\ref{eq:EOM_gen_scaled}). Of course, it is possible to numerically integrate equations~(\ref{eq:EOM_gen_scaled}), which is done here.

\begin{figure*}
\centering
\includegraphics[scale = 0.42, trim = 35mm 0mm 0mm 0mm]{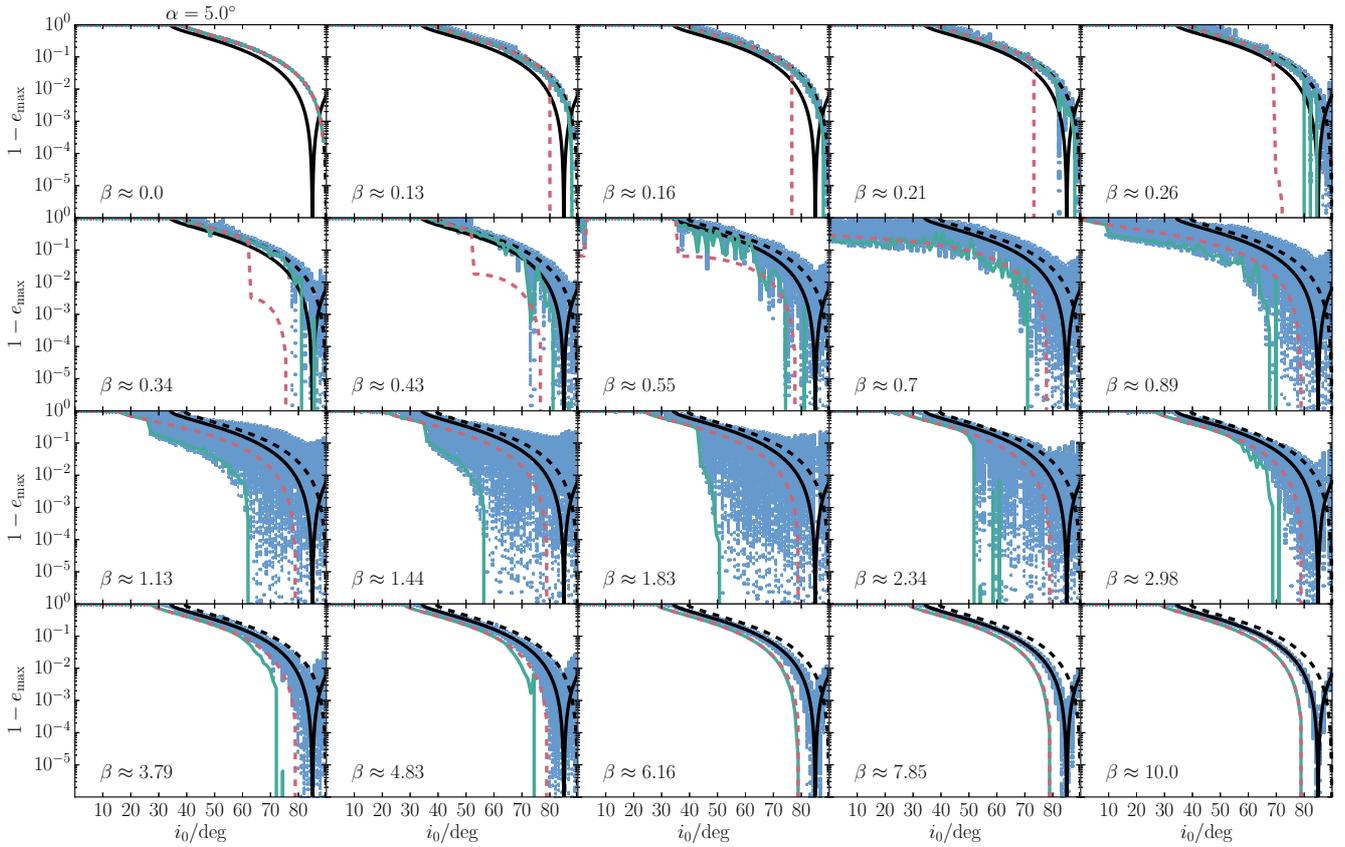}
\caption{\small Blue points: the (local) maximum eccentricities as a function of the initial inclination $i_0$ between $\ve{j}$ and $\unit{L}_\mathrm{out}$, determined by numerically solving equations~(\ref{eq:EOM_gen_scaled}) with $\alpha=5^\circ$. Each of the 20 panels corresponds to a different value of $\beta$ indicated therein; the number of bins in $i_0$ is 80. The black dashed lines show the canonical result (equation~\ref{eq:e_max_beta_zero}), which applies in the limit $\beta=0$. The solid black lines show equation~(\ref{eq:e_max_inf_beta}), which is valid in the limit $\beta\gg1$. The solid green lines show the maximum eccentricities computed from the numerically-determined maximum inclinations and the ad hoc equation~(\ref{eq:e_max_i_max_ad_hoc}) (see \S\,\ref{sect:num_gen:intermediate}). Similarly, the red dashed lines show the maximum eccentricities computed using equation~(\ref{eq:e_max_i_max_ad_hoc}) with $i_\mathrm{max}$ determined from the inclination model (see \S\,\ref{sect:incl_model:maxi}). }
\label{fig:fig3_emax_i_rel}
\end{figure*}

The initial conditions for the numerical integrations are as follows. The vectors $\unit{L}$, $\unit{L}_\mathrm{out}$ and $\unit{z}$ are initially all assumed to lie in the same plane; the latter plane is the $(x,z)$-plane. In addition, $\alpha$, the angle about which $\unit{L}_\mathrm{out}$ precesses around $\unit{z}$, is defined such that positive $\alpha$ corresponds to $\unit{L}_\mathrm{out}$ having a positive component along the $x$ axis. The initial mutual inclination between $\unit{L}$ and $\unit{L}_\mathrm{out}$ is denoted with $i_0$. The initial mutual inclination between $\unit{L}$ and $\unit{z}$ is $i_{z,0} = i_0 + \alpha$. The eccentricity vector $\ve{e}$ is assumed to be initially parallel to the $\unit{y}$ axis, and its initial magnitude is set to $e=0.01$. 

We carried out sets of 1600 integrations on a linear grid with $\beta \equiv \Omega_\mathrm{out} \, t_\mathrm{LK}$ running from 0 to 10, and the initial $i$, $i_0$, running from $0^\circ$ to $89^\circ$. Here, we set $\alpha = 5^\circ$ or $30^\circ$ for each set. The duration of each integration was $\Delta \tau = 1000$, i.e., corresponding to a physical time-span of $1000 \, t_\mathrm{LK}$, and approximately 1000 LK oscillations if $\unit{L}_\mathrm{out}$ were fixed. During the integrations, local maxima of $e$ were recorded, and the latter are shown as a function of $i_0$ in \F\,\ref{fig:fig3_emax_i_rel} for $\alpha=5^\circ$ (results for $\alpha=30^\circ$ are given in Appendix~\ref{app:alpha_30}). Each of the 20 panels corresponds to a different value of $\beta$ indicated therein; the number of bins in $i_0$ is 80. 

Below, we discuss the maximum eccentricity behaviour as a function of $i_0$ for three regimes of $\beta$: small $\beta$ (\S\,\ref{sect:num_gen:small}), large $\beta$ (\S\,\ref{sect:num_gen:large}) and intermediate $\beta$  (\S\,\ref{sect:num_gen:intermediate}). Short-range forces are discussed briefly in \S\,\ref{sect:num_gen:SRF}. In \S\,\ref{sect:incl_model}, we present a simplified model for the mutual inclination, and use this to approximately describe the behaviour in the regime of intermediate $\beta$.

\subsection{Small $\beta$}
\label{sect:num_gen:small}

In each panel of \F\,\ref{fig:fig3_emax_i_rel}, the black dashed curve shows the canonical expression
\begin{align}
\label{eq:e_max_beta_zero}
e_\mathrm{max} = \sqrt{1- \frac{5}{3} \cos^2(i_0)}, 
\end{align}
which applies to the `unperturbed' problem with $\beta =0$, and assuming initially $e=0$ and $\cos(i_0) < \sqrt{3/5}$. This result, which is described here in detail for further reference, can be obtained by dotting equation~(\ref{eq:EOM_gen_scaled:e}) with $\unit{e}$, giving the stationary points, i.e.,
\begin{align}
0 = \frac{\mathrm{d} e}{\mathrm{d} \tau} = - \frac{15}{4} \left ( \ve{e} \cdot \unit{L}_\mathrm{out} \right ) \left [ \unit{e} \cdot \left ( \ve{j} \times \unit{L}_\mathrm{out} \right ) \right ].
\end{align}
Assuming that at the stationary point $e\neq 0$, $e\neq 1$ and $\unit{e} \cdot \unit{L}_\mathrm{out} \neq 0$, this can be rewritten as the condition
\begin{align}
\label{eq:max_e_der1}
\left ( \unit{e} \cdot \unit{L}_\mathrm{out} \right )^2 = 1 - \left ( \unit{L} \cdot \unit{L}_\mathrm{out} \right )^2,
\end{align}
where we used a vector identity for the scalar product of two triple products. Using equation~(\ref{eq:max_e_der1}) to eliminate $\unit{e} \cdot \unit{L}_\mathrm{out}$, and using the LK constant to express $\unit{L} \cdot \unit{L}_\mathrm{out}$ in terms of the initial inclination, conservation of the Hamiltonian yields an algebraic equation for $e$ with the solution equation~(\ref{eq:e_max_beta_zero}) for the (maximum) stationary point. 

As shown in \F\,\ref{fig:fig3_emax_i_rel}, for the smallest non-zero values of $\beta$, equation~(\ref{eq:e_max_beta_zero}) still gives a good description as expected. For $\beta \gtrsim 0.2$ but still $\ll1$, noticeable deviations start to occur; in particular, large eccentricity excitations, much larger than based on equation~(\ref{eq:e_max_beta_zero}), are possible for inclinations near $90^\circ$.

\subsection{Large $\beta$}
\label{sect:num_gen:large}
If $\beta$ is large ($\beta \gg 1$), then $\unit{L}_\mathrm{out}$ is precessing rapidly around the $\unit{z}$ axis compared to the time-scale at which $\unit{L}$ evolves (i.e., on a time-scale on the order of $t_\mathrm{LK}$). Intuitively, one might expect in this case that the rapid precession of $\unit{L}_\mathrm{out}$ effectively means that $\unit{L}_\mathrm{out}$ is pointing along $\unit{z}$, although with a modified `effective length' depending on $\alpha$, the angle at which $\unit{L}_\mathrm{out}$ precesses around $\unit{z}$. 

In the $\beta \gg 1$ limit, we can average the Hamiltonian equation~(\ref{eq:H_gen_rot}) over a precessional cycle of $\unit{L}_\mathrm{out}$. To achieve this, we write
\begin{align}
\unit{L}_\mathrm{out} = \cos(\phi) \sin(\alpha) \, \unit{x} + \sin(\phi) \sin(\alpha) \, \unit{y} + \cos(\alpha) \, \unit{z},
\end{align}
where $\phi\in [0,2\pi)$ is the (rapidly changing) precessional phase angle. The precession average of a quantity $(...)$ is defined by
\begin{align}
\langle (...) \rangle \equiv \frac{1}{2\pi} \int_0^{2\pi} \mathrm{d} \phi \, (...).
\end{align}
A number of useful averages are
\begin{subequations}
\begin{align}
\label{eq:prec_av_3}
\left \langle \left [ \unit{e} \cdot \left ( \ve{j} \times \unit{L}_\mathrm{out} \right ) \right ] \right \rangle &= \left [ \unit{e} \cdot \left ( \ve{j} \times \unit{z} \right ) \right ] \cos(\alpha); \\
\label{eq:prec_av_4}
\left \langle \left ( \unit{e} \cdot \unit{L}_\mathrm{out} \right )^2 \right \rangle &= \frac{1}{4} \left [ 1 + \left ( \unit{e} \cdot \unit{z} \right )^2 - \cos(2\alpha) \left \{ 1 - 3 \left ( \unit{e} \cdot \unit{z} \right )^2 \right \} \right ]; \\
\label{eq:prec_av_5}
\left \langle \left ( \unit{L} \cdot \unit{L}_\mathrm{out} \right )^2 \right \rangle &= \frac{1}{4} \left [ 1 + \left ( \unit{L} \cdot \unit{z} \right )^2 - \cos(2\alpha) \left \{ 1 - 3 \left ( \unit{L} \cdot \unit{z} \right )^2 \right \} \right ].
\end{align}
\end{subequations}

Averaging equation~(\ref{eq:LK_constant_nonzero}) gives 
\begin{align}
\label{eq:LK_inf_beta}
\frac{\mathrm{d}}{\mathrm{d}t} \left ( \ve{j} \cdot \unit{z} \right) \approx 0.
\end{align}
In other words, in the limit $\beta\gg 1$, the LK constant, $\ve{j} \cdot \unit{L}_\mathrm{out}$, is replaced by $\ve{j} \cdot \unit{z}$. Owing to the existence of this conserved quantity, the maximum eccentricity can be obtained analytically, as described below. 

Following the same steps as in \S\,\ref{sect:num_gen:small}, the condition for a stationary eccentricity (still) reads $\left [ \unit{e} \cdot \left ( \ve{j} \times \unit{L}_\mathrm{out} \right ) \right ] = 0$. From equation~(\ref{eq:prec_av_3}), it follows that $\left [ \unit{e} \cdot \left ( \ve{j} \times \unit{z} \right ) \right ] = 0$, which implies $\left ( \unit{e} \cdot \unit{z} \right )^2 = 1 - \left ( \unit{L} \cdot \unit{z} \right )^2$. 

Next, we substitute these results into the Hamiltonian equation~(\ref{eq:H_gen_rot}). By equation~(\ref{eq:LK_inf_beta}), the last term $\propto \ve{j} \cdot \unit{z}$ in the Hamiltonian is constant and can therefore be omitted. Dropping the latter term, using $\left ( \unit{e} \cdot \unit{z} \right )^2 = 1 - \left ( \unit{L} \cdot \unit{z} \right )^2$ to eliminate $\unit{e} \cdot \unit{z}$ and equation~(\ref{eq:LK_inf_beta}) to relate $\unit{L} \cdot \unit{z}$ to the initial value, $(\unit{L}\cdot \unit{z})_0 = \cos(i_{z,0})$, and applying equations~(\ref{eq:prec_av_4}) and (\ref{eq:prec_av_5}), the Hamiltonian at the stationary eccentricity in the limit $\beta\gg1$ reads
\begin{align}
\label{eq:H_inf_beta}
\nonumber \overline{H}_\mathrm{gen,rot,\beta \gg 1,\mathrm{stat}} &= H_0 \left [ 1-6e^2 + \frac{15}{4} e^2 \left \{ 2 - \frac{ \cos^2(i_{z,0})}{1-e^2} - \cos(2 \alpha) \left ( -2 + 3 \frac{ \cos^2(i_{z,0})}{1-e^2}  \right ) \right \} \right. \\
&\quad \left. - \frac{3}{4} \left(1-e^2\right)  \left \{ 1 + \frac{ \cos^2(i_{z,0})}{1-e^2} - \cos(2 \alpha) \left ( 1 - 3 \frac{ \cos^2(i_{z,0})}{1-e^2}  \right ) \right \} \right ].
\end{align}
Equating~(\ref{eq:H_inf_beta}) to the initial averaged Hamiltonian with $e=0$,
\begin{align}
\nonumber \overline{H}_\mathrm{gen,rot,\beta \gg 1,\mathrm{init}} = H_0 \left [ 1 - \frac{3}{4} \left \{ 1 + \cos^2(i_{z,0}) - \cos(2\alpha) \left ( 1 - 3 \cos^2(i_{z,0}) \right ) \right \} \right ],
\end{align}
we can solve analytically for the maximum eccentricity as a function of $i_{z,0}$ and $\alpha$. The solution is
\begin{align}
\label{eq:e_max_inf_beta}
e_\mathrm{max} &= \sqrt{1 - \frac{5}{3} \cos^2(i_{z,0})},
\end{align}
which is, remarkably, independent of $\alpha$ (if expressed in terms of $i_{z,0}$). Equation~(\ref{eq:e_max_inf_beta}) is simply equation~(\ref{eq:e_max_beta_zero}) with $i_0$ replaced by $i_{z,0}$. In other words, in the limit of very rapid precession, one obtains the classical result for the maximum eccentricity with the initial mutual inclination now replaced by the initial inclination between $\unit{L}$ and the $\unit{z}$ axis.

In \F\,\ref{fig:fig3_emax_i_rel}, we plot equation~(\ref{eq:e_max_inf_beta}) with the solid black lines. The numerical integrations indeed agree well with equation~(\ref{eq:e_max_inf_beta}) for large $\beta$ ($\beta \gtrsim 6$). Note that the largest $e_\mathrm{max}$ does not occur at $i_0=90^\circ$, but at $i_0=90^\circ-5^\circ=85^\circ$, since $\alpha=5^\circ$. 

\begin{figure}
\centering
\includegraphics[scale = 0.42, trim = 35mm 0mm 0mm 0mm]{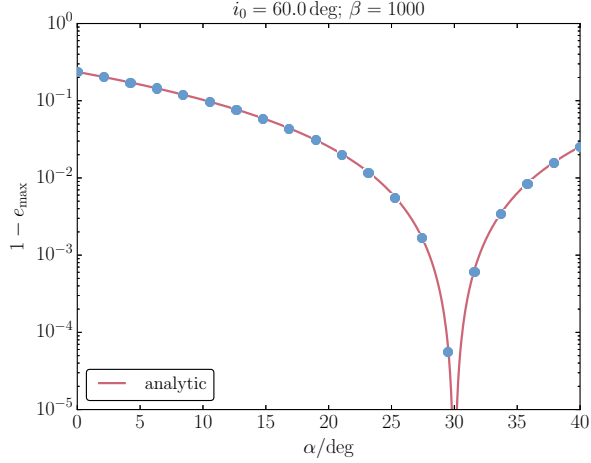}
\caption{\small Blue points: the maximum eccentricities as a function of $\alpha$ determined by numerically integrating the equations of motion. In this case, we set $i_0=60^\circ$ and $\beta=1000$. The solid red line shows the analytic result, equation~(\ref{eq:e_max_inf_beta}), which applies in the limit $\beta\gg 1$ (see \S\,\ref{sect:num_gen:large}). }
\label{fig:e_max_alpha}
\end{figure}

As an additional test of equation~(\ref{eq:e_max_inf_beta}), we show in \F\,\ref{fig:e_max_alpha} with blue points the maximum eccentricities as a function of $\alpha$ determined by numerically integrating the equations of motion, where we set $i_0=60^\circ$ and $\beta=1000$. The numerical results are in good agreement with equation~(\ref{eq:e_max_inf_beta}), shown with the solid red line.

\subsection{Intermediate $\beta$}
\label{sect:num_gen:intermediate}
In the intermediate-$\beta$ regime ($0.2 \lesssim \beta \lesssim 6$), $e_\mathrm{max}$ as a function of $i_0$ is much more complex. In particular, for $\beta$ around unity, extremely large eccentricity excitations are possible. Whereas in the limit $\beta=0$, $e_\mathrm{max}\rightarrow 1$ only for $i_0\rightarrow90^\circ$, if $\beta\sim1$, $e_\mathrm{max}\rightarrow 1$ for a much larger range of $i_0$ with $1-e_\mathrm{max}$ in the integrations reaching values lower than $10^{-6}$. The precise range of $i_0$ for extreme eccentricity excitation depends on $\beta$ (and $\alpha$; compare, Figures\,\ref{fig:fig3_emax_i_rel} and \ref{fig:fig3_emax_i_rel_30}). For lower $\beta$, $0.2 \lesssim \beta \lesssim 0.5$, and $\alpha=5^\circ$, there are specific inclinations $i_0$ for which $e_\mathrm{max}\rightarrow 1$, indicative of an overlap of resonances, and, therefore, chaos \citep{1979PhR....52..263C}. For larger $\beta$, $0.5 \lesssim \beta \lesssim 3$, there is a broad range of $i_0$ for which extreme eccentricities are reached. This range of local maximum eccentricities is manifested as a band in the $(i_0,1-e_\mathrm{max})$ plane. The upper boundary on $e_\mathrm{max}$ of this band appears to be smooth function of $i_0$. A (qualitative) understanding of this boundary is useful for practical applications, since it describes the largest possible eccentricity (i.e., over all local values of $e_\mathrm{max}$) for a given $i_0$. 

\begin{figure*}
\centering
\includegraphics[scale = 0.57, trim = 20mm 0mm 0mm 0mm]{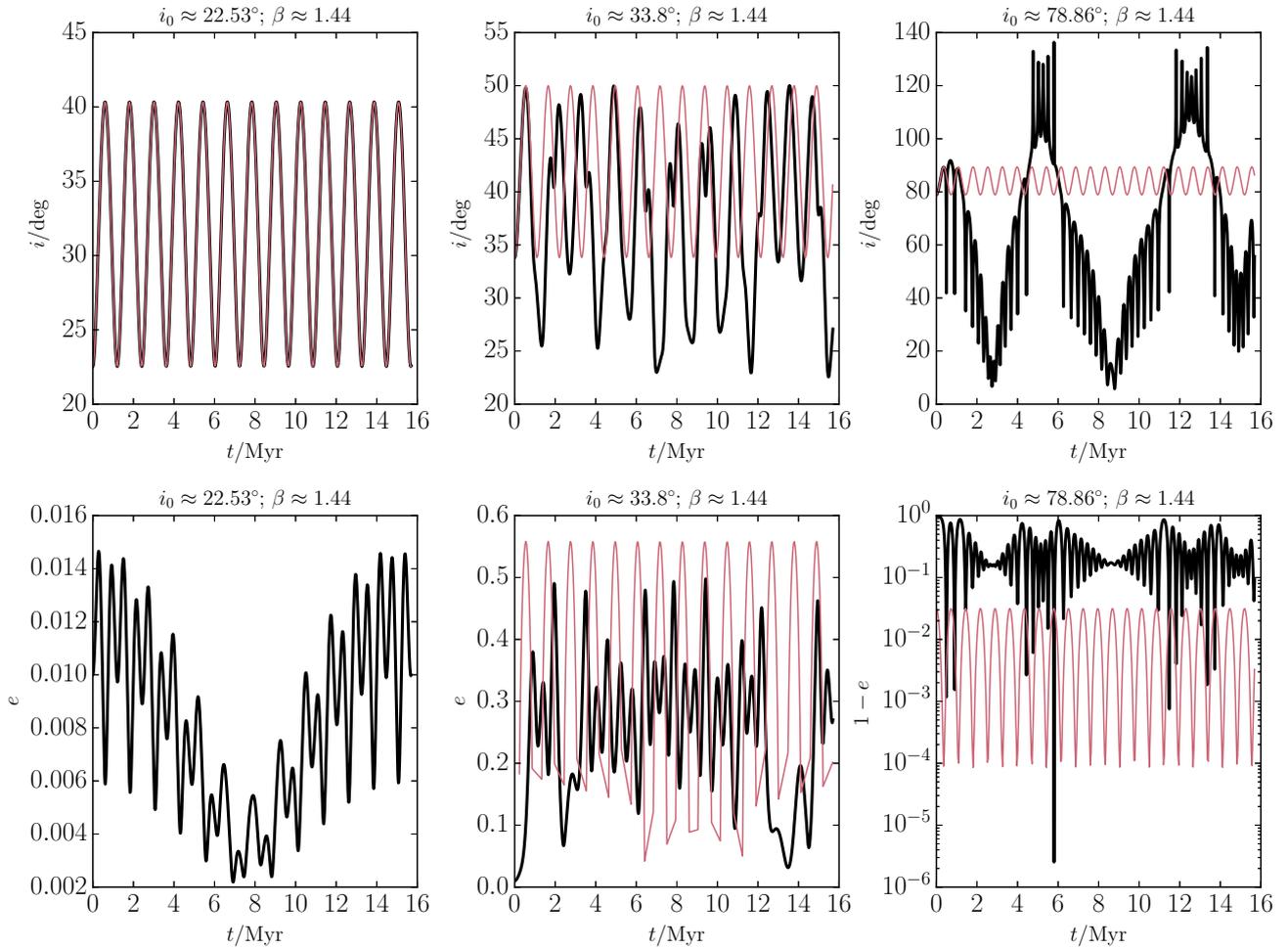}
\caption{\small A number of examples of the inclination and eccentricity evolution as a function of time, for a reduced time span compared to \F\,\ref{fig:fig3_emax_i_rel}. The initial inclination $i_0$ and the value of $\beta$ are indicated in the top of each panel. In all cases, $\alpha=5^\circ$. Black lines are according to the general model of \S\,\ref{sect:model}. Red lines are according to the inclination model (\S\,\ref{sect:incl_model}), for which the maximum eccentricities are computed from the ad hoc equation~(\ref{eq:e_max_i_max_ad_hoc}) (with the exception for the example in which $i_0 \approx 22.53^\circ$). }
\label{fig:example_test06_ex}
\end{figure*}

In the chaotic regime of intermediate $\beta$, the eccentricity generally evolves in a complicated way, with local maxima occurring at different values, and orbital flips occurring frequently. A number of examples of the inclination and eccentricity as a function of time for $\beta\approx 1.44$ and $\alpha=5^\circ$ are given in \F\,\ref{fig:example_test06_ex}. In particular, for $i_0\approx 34^\circ$, there is significant eccentricity excitation even though $i_0$ is less than the critical LK angle. For $i_0 \approx 79^\circ$, orbital flips occur, and they are associated with extremely high eccentricities. 

\begin{figure*}
\centering
\includegraphics[scale = 0.42, trim = 40mm 0mm 0mm 0mm]{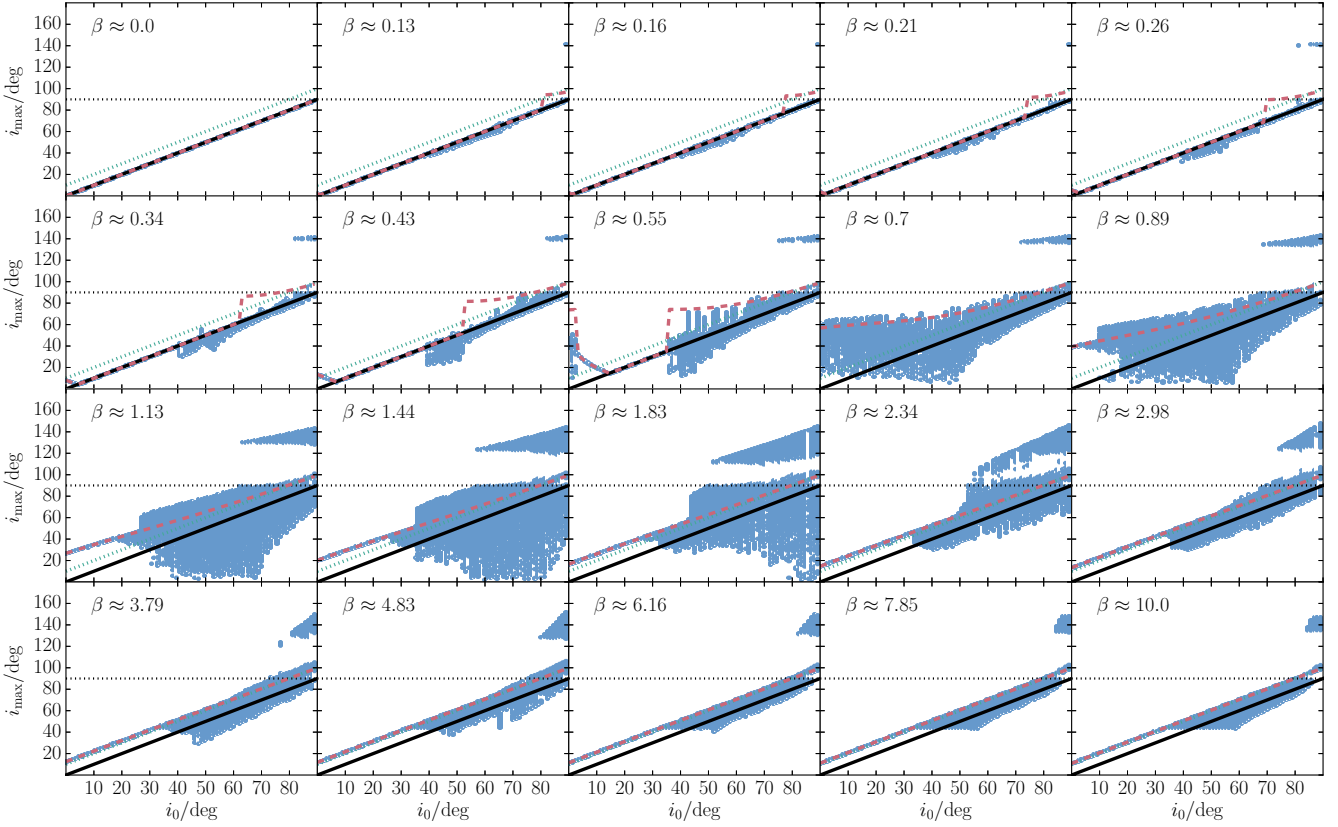}
\caption{\small Blue points: the (local) maximum inclinations as a function of $i_0$, determined numerically by solving equations~(\ref{eq:EOM_gen_scaled}) with $\alpha=5^\circ$. Each of the 20 panels corresponds to a different value of $\beta$ indicated therein; the number of bins in $i_0$ is 80. The black solid lines show $i_\mathrm{max}=i_0$, expected in the limit $\beta=0$. The green dotted lines show $i_\mathrm{max}=i_0+2\alpha$, expected in the limit $\beta\gg 1$ (\S\,\ref{sect:num_gen:intermediate}). The red dashed lines correspond to the inclination model (\S\,\ref{sect:incl_model:maxi}). }
\label{fig:fig1_imax_i_rel}
\end{figure*}

In order to understand the underlying mechanism for enhanced eccentricity excitation for a given $i_0$, we explore the connection between the maximum inclination, $i_\mathrm{max}$, and the maximum eccentricity. This is motivated by the notion that, if the mutual inclination is initially small, a large mutual inclination can be generated between $\unit{L}$ and $\unit{L}_\mathrm{out}$ because of secular angular-momentum evolution. Subsequently, the LK mechanism can be triggered if the generated mutual inclinations are sufficiently large. 

We determine from the numerical integrations the local maximum inclinations, which are shown as a function of $i_0$ for $\alpha=5^\circ$ in \F\,\ref{fig:fig1_imax_i_rel}. For $\beta\ll 1$, $i_\mathrm{max}\approx i_0$; the latter is shown with the black solid lines. For $\beta \gg 1$, $i_\mathrm{max} \approx i_0 + 2 \alpha$; the latter is shown with the green dotted lines. The large-$\beta$ behaviour can be explained by noting that if $\unit{L}_\mathrm{out}$ is precessing rapidly, the dynamics of the inner binary are the same as if $\unit{L}_\mathrm{out}$ were aligned along the $\unit{z}$ axis (see \S\,\ref{sect:num_gen:large}). However, as a consequence of the initial geometry, the maximum inclination now occurs when the nodal angle of $\unit{L}$ with respect to $\unit{L}_\mathrm{out}$ is $\pi$, such that the angle between $\unit{L}$ and $\unit{L}_\mathrm{out}$ is $i_0 + 2 \alpha$ .

In the intermediate $\beta$ regime, $i_\mathrm{max}$ can be much higher than either $i_0$ or $i_0+2\alpha$. For $0.6 \lesssim \beta \lesssim 2$, $i_\mathrm{max}$ is larger than $i_0+2\alpha$ for a large range of $i_0$. In addition, depending on $\beta$, above a critical angle of $i_0$ there is a jump in $i_\mathrm{max}$. For example, for $\beta \approx 0.89$, this jump occurs at $i_0 \approx 10^\circ$; for $\beta \approx 1.83$, it occurs at $i_0 \approx 45^\circ$. These jumps are also reflected in the maximum eccentricities (see \F\,\ref{fig:fig3_emax_i_rel}). 

For a range of values $\beta$, orbital flips occur, i.e., $i_\mathrm{max}>90^\circ$ and the orbital orientation switches from prograde to retrograde. Note that such flips are not possible for $\beta=0$ (we are assuming the quadrupole-order approximation; flips do occur for $\beta=0$ at the octupole order, e.g., \citealt{2011Natur.473..187N}). For $0.21\lesssim \beta \lesssim 0.43$, flips only occur if $i_0$ is close to $90^\circ$. As expected, these flips are associated with high eccentricities, which can be seen when compared to \F\,\ref{fig:fig3_emax_i_rel}. For $\beta$ closer to unity, the parameter space for flips is much larger, and a large region of retrograde orbits is populated with $i_0$ as low as $\approx 50^\circ$. 

The maximum inclinations are generally strongly related to the maximum eccentricities. In \F\,\ref{fig:fig3_emax_i_rel}, we show with solid green lines the ad hoc expression
\begin{align}
\label{eq:e_max_i_max_ad_hoc}
e_\mathrm{max} = \sqrt{1-\frac{5}{3} \cos^2(i_\mathrm{max})},
\end{align}
where $i_\mathrm{max}$ is the largest local maximum inclination determined from the numerical integrations. If $i_\mathrm{max}\geq90^\circ$, we set $e_\mathrm{max}=1$, which is motivated by the expectation that if retrograde orbits are attained from initially prograde orbits, then $i=90^\circ$ at certain times in the evolution; moreover, orbital flips are typically associated with $e\rightarrow 1$. 

Except in the limit of large $\beta$, the ad hoc expression equation~(\ref{eq:e_max_i_max_ad_hoc}) captures the maximum eccentricities. In particular, the shape of the maximum eccentricity envelopes is captured by equation~(\ref{eq:e_max_i_max_ad_hoc}); the same applies for the peaks or spikes, occurring for $0.26\lesssim \beta \lesssim 0.55$. This agreement supports the notion that high mutual inclinations, arising from precession of $\unit{L}_\mathrm{out}$, can drive high eccentricities through the LK mechanism. Therefore, in order to understand the dependence of $e_\mathrm{max}$ on $i_0$, it is useful to understand how $i_\mathrm{max}$ depends on $i_0$. For this purpose, we discuss below, in \S\,\ref{sect:incl_model}, a simplified model that describes the inclination evolution only. This simplified model is more amenable to analytic treatment, and is therefore useful to gain more insight.

\subsection{Short-range forces}
\label{sect:num_gen:SRF}
We briefly discuss how the above results are modified if short-range forces are included. It is well known that short-range forces due to general relativity and tidal/ rotational bulges of the star/planet tend to suppress eccentricity excitation or limit the maximum eccentricity that can be achieved in LK oscillations (e.g., \citealt{2003ApJ...589..605W}; see \citealt{2007ApJ...669.1298F,2015MNRAS.447..747L} for analytic calculations). Here, we include general relativistic precession at the first post-Newtonian order. The additional precession breaks the scale invariance of the system; therefore, the absolute physical scales need to be specified. The latter are set as follows: the inner and outer semimajor axes and eccentricities are set to 1 and 100 AU, and 0.01 and 0.1, respectively. The assumed masses are $m_0=m_2 = 1 \, \mathrm{M}_\odot$, and $m_1 = 1 \, M_\mathrm{J}$. With these choices, the characteristic LK time-scale is $t_\mathrm{LK} \approx 0.16 \, \mathrm{Myr}$, whereas the relativistic precession time-scale is 
\begin{align}
t_\mathrm{1PN} = \frac{1}{3} P_\mathrm{orb,1} \left (1-e^2 \right ) \frac{a}{r_\mathrm{g}} \approx 33.7 \, \left (1-e^2 \right ) \, \mathrm{Myr},
\end{align}
where $P_\mathrm{orb,1}$ is the inner orbital period, and $r_\mathrm{g} = G(m_0+m_1)/c^2$ is the gravitational radius. Note that for the assumed masses and semimajor axes, tidal effects play an important role (see \citealt{2015MNRAS.447..747L,2016MNRAS.456.3671A}), so our integrations in this section are for illustrative purposes only.

\begin{figure*}
\centering
\includegraphics[scale = 0.42, trim = 35mm 0mm 0mm 0mm]{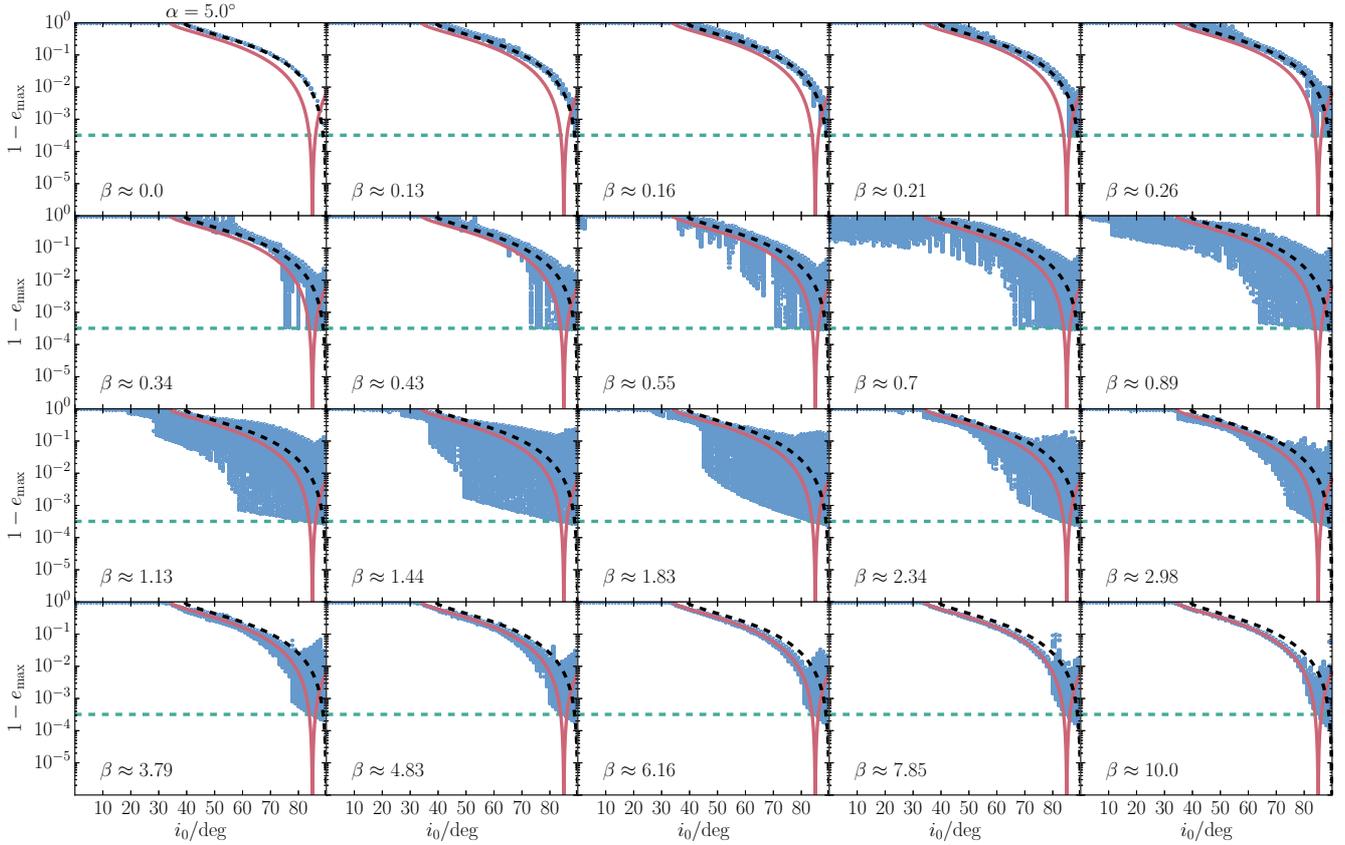}
\caption{\small Similar to \F\,\ref{fig:fig3_emax_i_rel}, now including general relativistic precession in the inner orbit (\S\,\ref{sect:num_gen:SRF}). The horizontal green dashed lines show the smallest $1-e_\mathrm{max}$ reached if $\beta=0$. }
\label{fig:fig1_emax_i_rel_test06_PN}
\end{figure*}

In \F\,\ref{fig:fig1_emax_i_rel_test06_PN}, we show the maximum eccentricities as a function of $i_0$ for $\alpha=5^\circ$ similar to \F\,\ref{fig:fig3_emax_i_rel}, now with the addition of general relativistic precession. The maximum eccentricities are generally the same compared to the situation without short-range forces, except when $1-e_\mathrm{max}$ is less than $\sim 1-10^{-3}$, for which there is a limit on $e_\mathrm{max}$. The horizontal green dashed lines show the smallest $1-e_\mathrm{max}$ reached for $\beta=0$ if $i_0 =90^\circ$. The latter value can be computed semi-analytically at the quadrupole order (e.g., \citealt{2002ApJ...578..775B}). The maximum eccentricity in the case $\beta=0$ is approximately the same compared to the case $\beta\neq 0$. Interestingly, the maximum eccentricities for $\beta \neq 0$ are (marginally) larger compared to the maximum $\beta=0$ value for $\beta \gtrsim 1$. 

In summary, although the maximum eccentricities are limited by general relativistic precession, there is still significant enhancement of the eccentricity compared to $\beta=0$ for small inclinations.

\section{A simplified model for the mutual inclination}
\label{sect:incl_model}
In order to better understand the dynamics underlying the extreme eccentricity excitation observed in \S\,\ref{sect:num_gen} in the intermediate $\beta$ regime, we explore a simplified model, henceforth referred to as the `inclination model', in which we consider evolution of the angular-momentum vectors only, and disregard changes in the eccentricity vectors. This model is motivated by the observation made in \S\,\ref{sect:num_gen:intermediate} that the qualitative behaviour of $e_\mathrm{max}$ as a function of $i_0$ can be explained by determining the maximum inclinations from the numerical integrations, and applying to these the ad hoc expression equation~(\ref{eq:e_max_i_max_ad_hoc}). This indicates that if high inclinations can be attained due to interaction with the precessing $\unit{L}_\mathrm{out}$, high eccentricities are triggered by the LK mechanism. Of course, our model is only an approximation because we consider the inclination and eccentricity evolution to be decoupled, i.e., the maximum inclination is driven by the interaction with $\unit{L}_\mathrm{out}$ and the maximum eccentricity is driven by $i_\mathrm{max}$, the latter neglecting the fact that $\unit{L}_\mathrm{out}$ is precessing. We note that the inclination model is mathematically very similar to the Hamiltonian model of \citet{2017AJ....153...42L}.

\subsection{Model description}
\label{sect:incl_model:descr}
The inclination model is obtained by setting $e=0$ in the model of \S\,\ref{sect:model}, i.e., 
\begin{subequations}
\label{eq:EOM_simple_scaled}
\begin{align}
\label{eq:EOM_simple_scaled:j}
\frac{\mathrm{d} \unit{L}}{\mathrm{d} \tau} &= \frac{3}{4} \left ( \unit{L} \cdot \unit{L}_\mathrm{out} \right ) \left ( \unit{L} \times \unit{L}_\mathrm{out} \right ); \\
\label{eq:EOM_simple_scaled:L_out}
\frac{\mathrm{d} \unit{L}_\mathrm{out}}{\mathrm{d} \tau} &= - \beta \left( \unit{z} \times \unit{L}_\mathrm{out} \right ).
\end{align}
\end{subequations}
The Hamiltonian is given by (see equation~\ref{eq:H_gen_rot})
\begin{align}
\label{eq:H_simple_rot}
H &= H_0 \left [ 1 - 3 \left ( \unit{L} \cdot \unit{L}_\mathrm{out} \right )^2 + 8 \beta  \left ( \unit{L} \cdot \unit{z} \right ) \right ].
\end{align}
To examine the evolution of $i$, the angle between $\unit{L}$ and $\unit{L}_\mathrm{out}$, we set up a rotating frame $(x',y'z')$ such that $\unit{z}'=\unit{L}_\mathrm{out}$ and $\unit{z}= -\sin(\alpha) \, \unit{x}'+\cos(\alpha)\,\unit{z}'$. Let $\Omega$ be the nodal angle [measured from the $x'$-axis in the $(x',y'$) plane] of the inner binary, such that $\unit{L}=\cos(\Omega) \sin(i) \, \unit{x}' + \sin(\Omega)\sin(i) \, \unit{y}' + \cos(i) \, \unit{z}'$. Then, equation~(\ref{eq:H_simple_rot}) reduces to
\begin{align}
\label{eq:H_circ_model}
H = H_0 \left [ 1 - 3 \cos^2 (i) + 8 \beta \left \{ \cos(\alpha) \cos(i) - \sin(\alpha) \cos(\Omega) \sin(i)  \right \} \right ].
\end{align}
Since $L \cos(i)$ and $\Omega$ are conjugate canonical variables (note that $L$ is related to $H_0$ according to $L=8 H_0 t_\mathrm{LK}$), the equations of motion for $(i,\Omega)$ are
\begin{subequations}
\label{eq:EOM_circ_model}
\begin{align}
\label{eq:EOM_circ_model:i}
\frac{\mathrm{d} i}{\mathrm{d} \tau} &= \beta \sin(\alpha) \sin(\Omega); \\
\label{eq:EOM_circ_model:Omega}
\frac{\mathrm{d} \Omega}{\mathrm{d} \tau} &= -\frac{3}{4} \cos(i) + \beta \left [ \cos(\alpha) + \sin(\alpha) \cos(\Omega) \cot(i) \right ].
\end{align}
\end{subequations}

\subsection{Maximum inclinations}
\label{sect:incl_model:maxi}
In \F\,\ref{fig:fig1_imax_i_rel}, the maximum inclinations obtained by numerically solving the equations of motion~(\ref{eq:EOM_circ_model}) are shown with the red dashed lines (below, in \S\,\ref{sect:incl_model:prop}, we show how these maximum inclinations can be computed semi-analytically). The inclination model captures several behaviours of $i_\mathrm{max}$.
\begin{enumerate}
\item The limit $i_\mathrm{max}= i_0$ for small $\beta$ and $i_\mathrm{max}=i_0+2\alpha$ for large $\beta$.
\item The enhanced $i_\mathrm{max}$ relative to $i_\mathrm{max}=i_0+2\alpha$ in the intermediate $\beta$ regime.
\item The increase of $i_\mathrm{max}$ for {\it decreasing} $i_0$ for $0.13 \lesssim \beta \lesssim 0.55$.
\item The presence of retrograde orbits for $i_0$ close to $90^\circ$; these flips are represented in \F\,\ref{fig:fig1_imax_i_rel} in the general model of \S\,\ref{sect:model} by the points near $140^\circ$, because the LK mechanism drives the inclination to $\approx 140^\circ$, which, in the integrations, is recorded as the maximum inclination.
\end{enumerate}
In \F\,\ref{fig:fig3_emax_i_rel}, the maximum eccentricities according to the ad hoc equation~(\ref{eq:e_max_i_max_ad_hoc}) computed with the maximum inclinations from the inclination model are shown with the red dashed lines (again, we set $e_\mathrm{max}=1$ if $i_\mathrm{max}>90^\circ$). Generally, the inclination model together with equation~(\ref{eq:e_max_i_max_ad_hoc}) captures the general trends of $e_\mathrm{max}$ with $\beta$. In particular, the envelopes for $0.55 \lesssim \beta \lesssim 0.7$ and $2.98 \lesssim \beta \lesssim 4.83$ are reproduced. Some of the features of high eccentricity at $0.13 \lesssim \beta \lesssim 0.34$ are not consistent with the general model, and the maximum eccentricity envelope is not captured for $1.13\lesssim \beta \lesssim 2.34$. The latter can be expected: the inclination model does not reproduce the additional enhancement in $i_\mathrm{max}$ observed in the generalized model, and the $e_\mathrm{max}$-envelope appears to be associated with this enhancement. Evidently, the extra enhancement in eccentricity (compare the red dashed and green lines in \F\,\ref{fig:fig3_emax_i_rel}) is due to a coupling between inclination induced in the inclination model, and the LK mechanism that becomes active at high inclinations. 

The inclinations from the inclination model and the implied maximum eccentricities through equation~(\ref{eq:e_max_i_max_ad_hoc}) are shown with the red lines in the examples of \F\,\ref{fig:example_test06_ex}. For $i_0\approx 23^\circ$, the inclination as a function of time is reproduced by the inclination model. For $i_0\approx 34^\circ$, the inclination model still captures some of the features of the inclination. In particular, the maximum inclination matches the value for the general model, and the implied eccentricities are similar.

\subsection{Properties of the model}
\label{sect:incl_model:prop}

\begin{figure*}
\centering
\includegraphics[scale = 0.45, trim = 10mm 0mm 0mm 0mm]{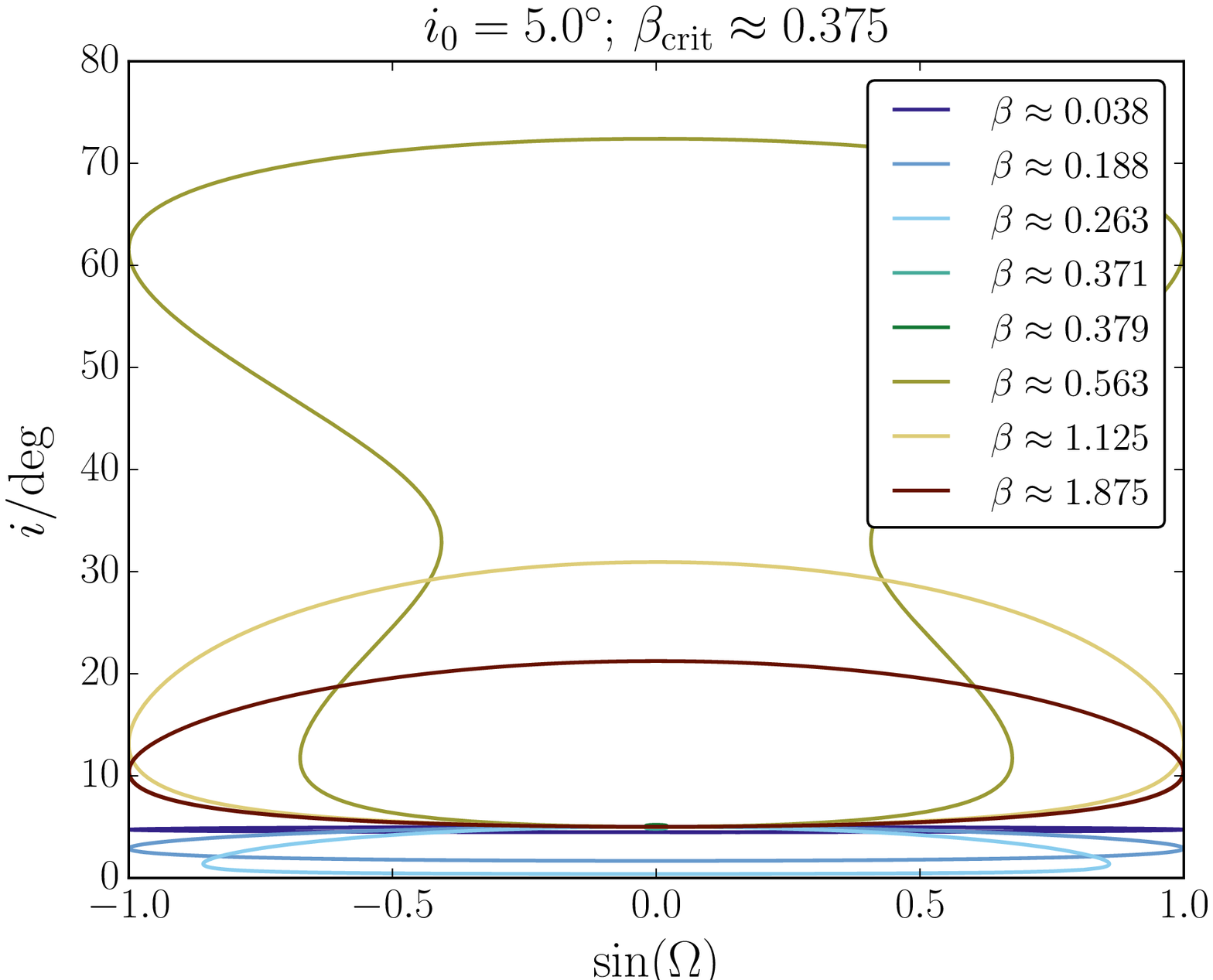}
\includegraphics[scale = 0.45, trim = 10mm 0mm 0mm 0mm]{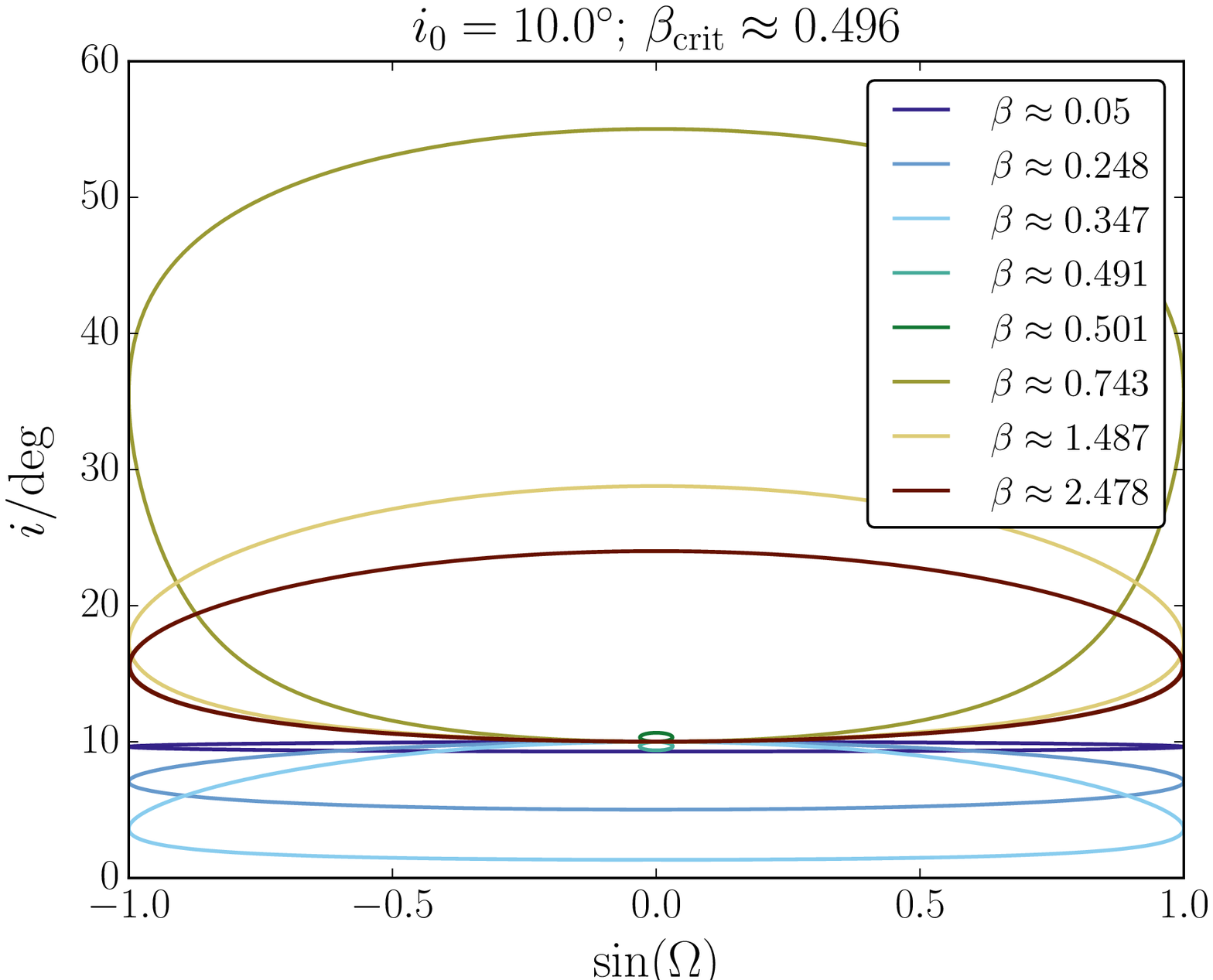}
\includegraphics[scale = 0.45, trim = 10mm 0mm 0mm 0mm]{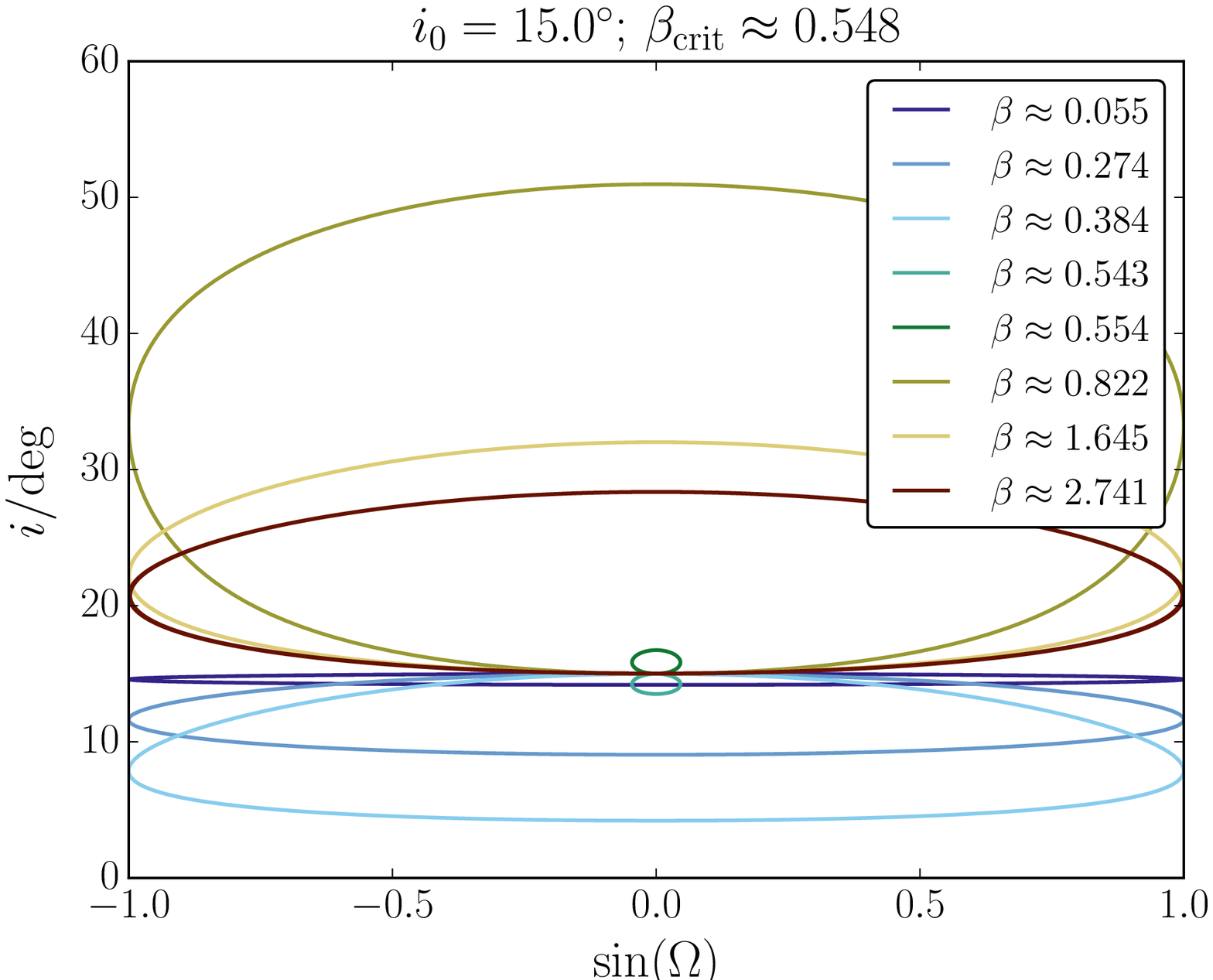}
\includegraphics[scale = 0.45, trim = 10mm 0mm 0mm 0mm]{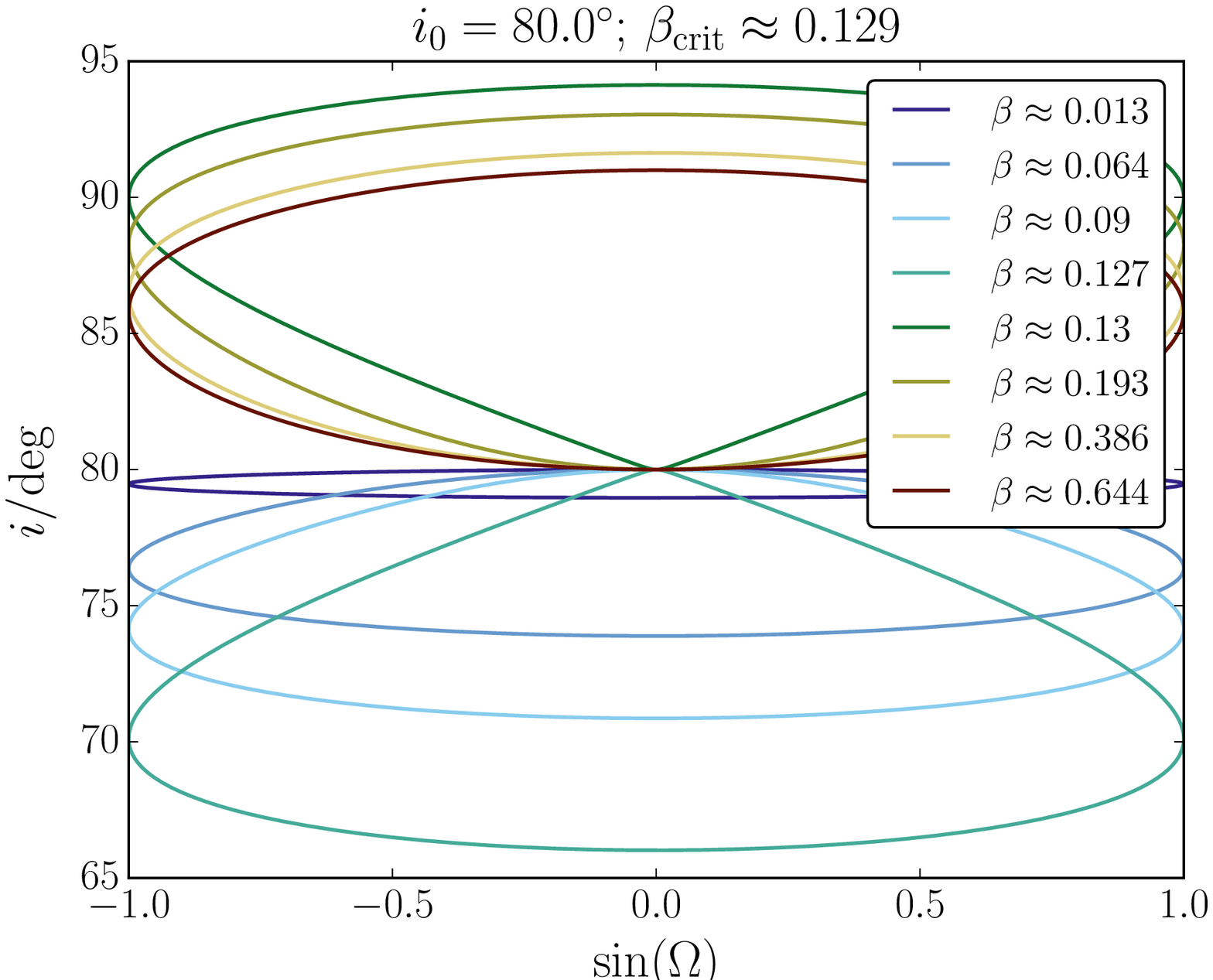}
\caption{ Phase-space curves in the $(\sin\Omega,i)$ plane according to the inclination model (see equation~\ref{eq:H_circ_model}), for different values of $\beta$ (indicated in the legends). Each panel corresponds to a different $i_0$ indicated in the top; $\alpha=5^\circ$ in all cases. Also indicated in the top is the critical value of $\beta$, $\beta_\mathrm{crit}$, for which the phase-space curves `flip' over (see \S\,\ref{sect:incl_model:prop} and equation~\ref{eq:beta_crit}). }
\label{fig:Omega_vs_i}
\end{figure*}

The maximum inclinations can be computed semi-analytically by using the Hamiltonian, equation~(\ref{eq:H_circ_model}). In \F\,\ref{fig:Omega_vs_i}, we show phase-space curves in the $(\sin\Omega,i)$ plane for different values of $\beta$ (with $\alpha=5^\circ$ fixed). Each panel corresponds to a different $i_0$. The following properties are revealed:
\begin{enumerate}
\item The inclination $i$ is always stationary at $\Omega=0$. This is also clear from equation~(\ref{eq:EOM_circ_model:i}). 
\item For small $\beta$ and $i_0$, $\Omega$ circulates, and the maximum inclination is $i_\mathrm{max}=i_0$, with the other stationary inclination $i_\mathrm{min}<i_0$. 
\item At a critical value of $\beta$, $\beta_\mathrm{crit}$, the curves `flip' over: the minimum inclination is now $i_0$, and the maximum inclination is $i_\mathrm{max}>i_0$ (note that this flip phenomenon is not an orbital flip). The value of $\beta_\mathrm{crit}$ depends on $i_0$.
\item For $\beta$ just above $\beta_\mathrm{crit}$, $i_\mathrm{max}$ increases with increasing $\beta$. As $\beta$ increases further, $i_\mathrm{max}$ decreases again.
\end{enumerate}

Since the stationary inclinations (including the maximum inclinations) occur at $\Omega=\pi$, $i_\mathrm{max}$ can be found from energy conservation (equation~\ref{eq:H_circ_model}),
\begin{align}
\label{eq:i_max_circ_model}
- 3 \cos^2 (i_0) + 8 \beta \left [ \cos(\alpha) \cos(i_0) - \sin(\alpha) \sin(i_0)  \right ] = - 3 \cos^2 (i) + 8 \beta \left [ \cos(\alpha) \cos(i) - \sin(\alpha) \sin(i)  \right ],
\end{align}
where we assumed that, initially, $\Omega=0$ (this is the case in all our numerical integrations). 

As can be seen from \F\,\ref{fig:Omega_vs_i} (particularly for small $i_0$), as $\beta$ increases from zero, at some point $\Omega$ no longer circulates but librates between two critical values of $\Omega$. At the critical $\beta$, the width of libration vanishes, i.e., $\Omega=0$ is constant, and $i=i_0$. Using the equations of motion (equation~\ref{eq:EOM_circ_model:Omega}), this implies that $\mathrm{d} \Omega/\mathrm{d} \tau=0$ for $\beta_\mathrm{crit}$, giving
\begin{align}
\label{eq:beta_crit}
\beta_\mathrm{crit} = \frac{3}{4} \frac{\cos (i_0)}{\cos(\alpha) + \sin(\alpha) \cot(i_0)}.
\end{align}
The value of $\beta_\mathrm{crit}$ is indicated in the top of each panel of \F\,\ref{fig:Omega_vs_i}. 

\begin{figure*}
\centering
\includegraphics[scale = 0.45, trim = 10mm 0mm 0mm 0mm]{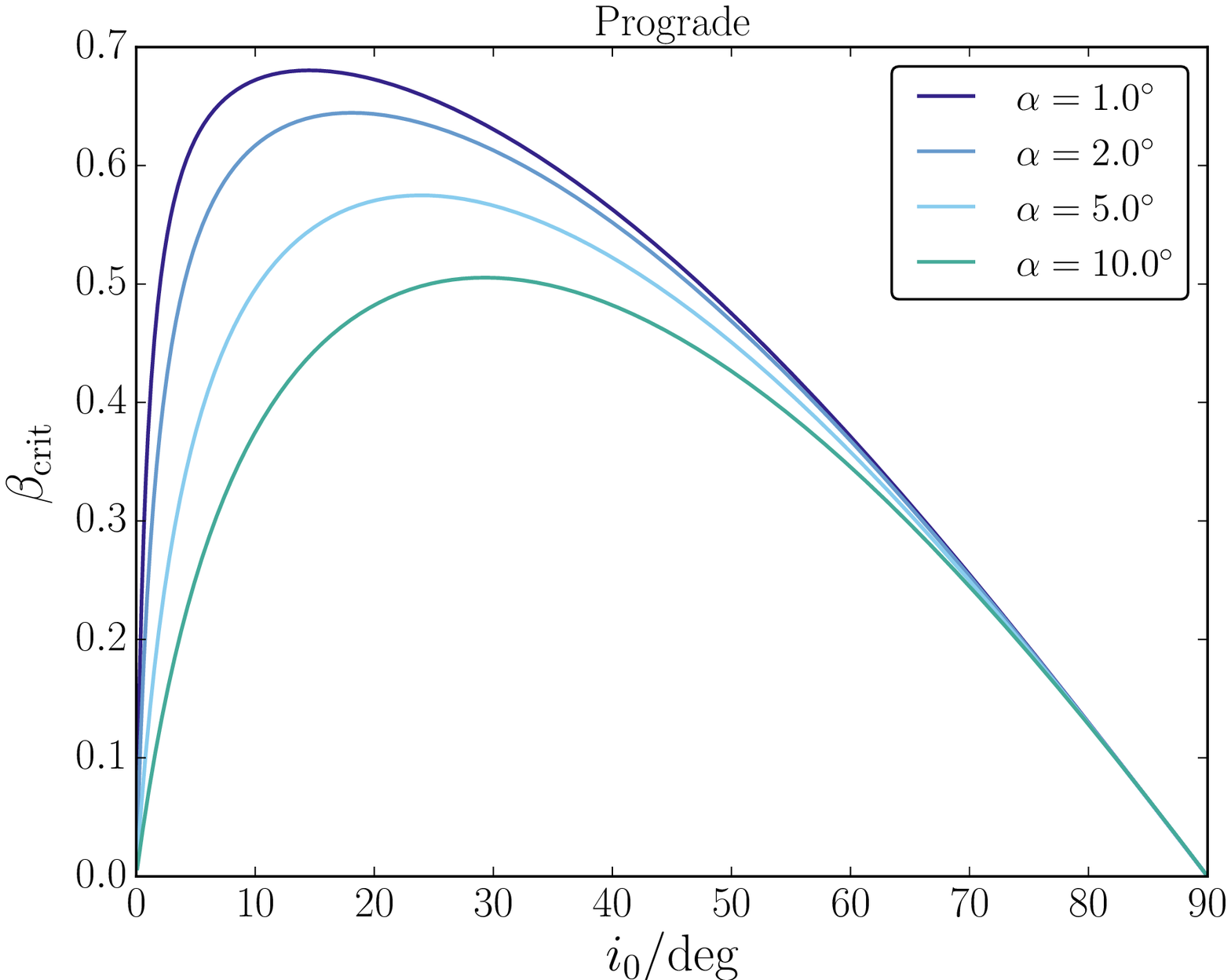}
\includegraphics[scale = 0.45, trim = 10mm 0mm 0mm 0mm]{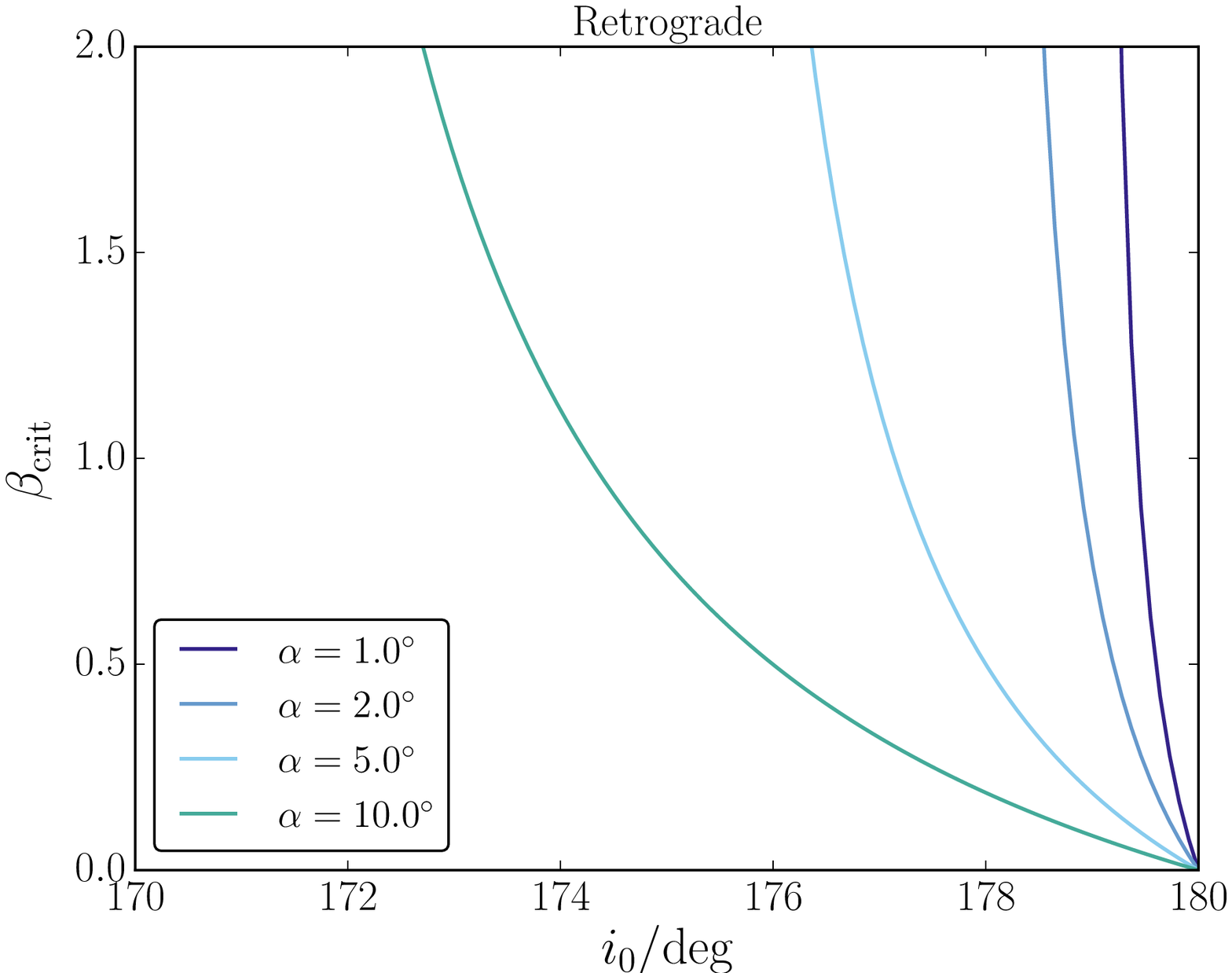}
\caption{The critical value of $\beta$, $\beta_\mathrm{crit}$ (equation~\ref{eq:beta_crit}), for which phase-space curve flips occur, as a function of $i_0$. Different colours correspond to different values of $\alpha$ (indicated in the legends). We distinguish between prograde ($i_0<90^\circ$; left-hand panel) and retrograde ($i_0>90^\circ$; right-hand panel) orbits. }
\label{fig:beta_crit}
\end{figure*}

In \F\,\ref{fig:beta_crit}, we plot equation~(\ref{eq:beta_crit}) as a function of $i_0$, for different values of $\alpha$. We distinguish between prograde ($i_0<90^\circ$; left-hand panel) and retrograde ($i_0>90^\circ$; right-hand panel) orbits. For $i_0>90^\circ$, $\beta_\mathrm{crit}$ becomes negative, i.e., the phase-space curve flips no longer occur. Depending on $\alpha$, $\beta_\mathrm{crit}$ becomes positive again for sufficiently retrograde orbits, showing that the flip phenomenon returns. 

\begin{figure}
\centering
\includegraphics[scale = 0.45, trim = 10mm 0mm 0mm 0mm]{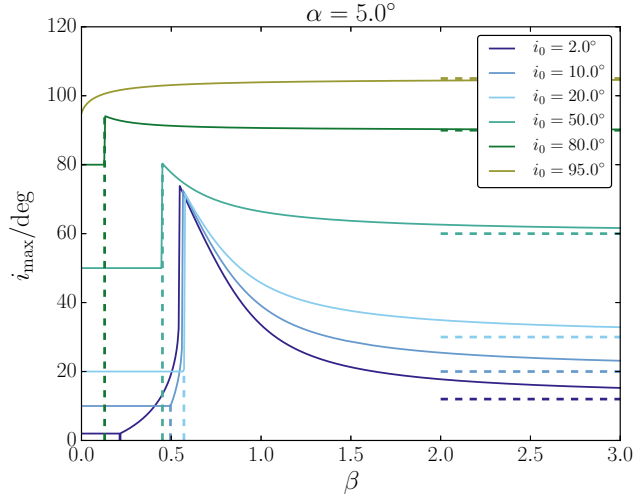}
\caption{ The maximum inclination as a function of $\beta$ according to the inclination model. Different colours correspond to different initial inclinations; in all cases, $\alpha=5^\circ$. The vertical dashed lines show the corresponding value of $\beta_\mathrm{crit}$ (equation~\ref{eq:beta_crit}). The horizontal dashed lines show the asymptotic limit $i_\mathrm{max}=i_0+2\alpha$.}
\label{fig:i_e_max_beta}
\end{figure}

In \F\,\ref{fig:i_e_max_beta}, we show the maximum inclination as a function of $\beta$. The values of $\beta_\mathrm{crit}$ corresponding to each $i_0$ (and the chosen $\alpha$) are shown with the vertical dashed lines. This figure illustrates how much the inclination can increase from the initial value $i_0$, provided that $\beta$ lies within a certain range. For example, even if only $i_0=2^\circ$, $i_\mathrm{max}$ can reach values of $\approx 70^\circ$ if $\beta$ is near 0.5. The minimum value of $\beta$ for which large inclinations are attained is approximated by $\beta_\mathrm{crit}$, unless $i_0$ is small (i.e., $\lesssim 10^\circ$). After rapidly reaching the peak value of $i_\mathrm{max}$ as $\beta$ increases beyond $\beta_\mathrm{crit}$, $i_\mathrm{max}$ steadily decreases (unless the orbit is initially retrograde). In the limit of large $\beta$, the maximum inclination reaches the expected limit $i_\mathrm{max} = i_0 + 2\alpha$, indicated in \F\,\ref{fig:i_e_max_beta} with the horizontal dashed lines. 

In summary, the inclination model shows that high mutual inclinations can be induced, even if $i_0$ is small. The degree of inclination enhancement depends sensitively on $\beta$: there is a range in $\beta$ for which $i_\mathrm{max}$ is large which depends on $i_0$ (and $\alpha$).

\section{Discussion}
\label{sect:discussion}

\subsection{Applicability of the inclination model}
\label{sect:discussion:incl}
We have shown that the inclination model (Section\,\ref{sect:incl_model}) qualitatively describes how high mutual inclinations can be achieved due to the nodal precession of the outer orbit, even for small initial mutual inclinations. We emphasize that the model, when combined with the ad hoc equation~(\ref{eq:e_max_i_max_ad_hoc}), does not quantitatively describe the intermediate-$\beta$ regime; in particular, the envelope of $e_\mathrm{max}$ with respect to $i_0$ in \F\,\ref{fig:fig3_emax_i_rel} is not as `deep' compared to the numerical integrations of the equations of motion. Therefore, the inclination model should not be used for accurate predictions of $e_\mathrm{max}$ in the intermediate-$\beta$ regime. However, it can be used to estimate the minimum $i_\mathrm{max}$ for a given $i_0$ and $\beta$, thereby giving an indication whether or not high eccentricities are to be expected. 

\subsection{Applications to astrophysical systems}
\label{sect:discussion:appl}
Amongst the astrophysical implications of the enhanced eccentricity oscillations discussed in this paper is the formation of HJs in stellar triple systems. We refer to section 3 of \citet{2017MNRAS.466.4107H} for an example of the `2+2' configuration, in which an HJ is formed through high-eccentricity migration in a stellar triple. In this example, migration is efficient, even though the initial mutual inclination between the proto-HJ and the orbit of the outer stars (i.e., orbit 3) is not large ($\approx 57^\circ$), and not high enough to induce migration in the case of a stellar binary (i.e., if orbit 2 were replaced by a point mass). In this example system, $\beta \approx 0.68$ showing that the system is in the regime where high mutual inclinations can be induced, driving high-eccentricity secular oscillations and tidal migration.

For reference, we give the explicit expressions for $\beta$ in terms of physical parameters\footnote{Equation~(\ref{eq:beta_to_phys_2p2}) is the same as equations~10 of \citet{2016MNRAS.459.2827H}, and 1 of \citet{2017MNRAS.466.4107H}, modulo a factor of $(3/4) \cos(i_{23,\mathrm{init}})$.  Equation~(\ref{eq:beta_to_phys_3p1}) is the same as equations~13 of \citet{2015MNRAS.449.4221H}, 9 of \citet{2016MNRAS.459.2827H}, and 3 of \citet{2017MNRAS.466.4107H}, again modulo a factor of $(3/4) \cos(i_{23,\mathrm{init}})$. }. For `2+2' systems,
\begin{align}
\label{eq:beta_to_phys_2p2}
\beta_{2+2} = \frac{3}{4} \cos(i_{23,\mathrm{init}}) \, \left ( \frac{a_2}{a_1} \right)^{3/2} \left ( \frac{m_0+m_1}{m_2+m_3} \right )^{3/2}.
\end{align}
Note that $a_3$ does not enter in this expression; it drops out in the ratio of the LK time-scales of orbit pairs (1,3) and (2,3). Of course, $a_3$ does set the absolute time-scale on which the secular evolution in the system occurs. For `3+1' systems,
\begin{align}
\label{eq:beta_to_phys_3p1}
\beta_{3+1} = \frac{3}{4} \cos(i_{23,\mathrm{init}}) \, \left ( \frac{a_2^3}{a_1 a_3^2} \right)^{3/2} \left ( \frac{m_0+m_1}{m_0+m_1 + m_2} \right )^{1/2} \frac{m_3}{m_2} \left ( \frac{1-e_2^2}{1-e_3^2} \right)^{3/2}.
\end{align}
Note that in this paper, we assumed $e_2=0$.

\section{Conclusions}
\label{sect:conclusions}
We have studied in detail the secular dynamics of hierarchical quadruple systems in either the `2+2' or `3+1' configurations (\F\,\ref{fig:configurations}). The `2+2' configuration applies, for example, to a planet orbiting a star, which is orbited by a more distant stellar binary (orbit 2), in a relatively wide orbit (orbit 3). An example of the `3+1' configuration is a planet orbiting a star, which is orbited by two more distant objects (one more distant than the other, in hierarchical orbits); we assume that the eccentricity of the orbit of the first distant object is zero throughout the evolution (i.e., the outer objects are inclined by no more than $\approx 40^\circ$ such that the LK mechanism does not operate). In particular, the `2+2' configuration applies to HJs in stellar triples \citep{2017MNRAS.466.4107H}, and the `3+1' configuration applies to hypothetical planetary companions to HJs in stellar binaries \citep{2017ApJ...835L..24H}. We have formulated the secular equations of motion for both configurations in one generalized model. In this model, the problem has been reduced to a hierarchical {\it three-body} problem with the perturbed outer orbital angular-momentum axis $\unit{L}_\mathrm{out}$ precessing around a fixed axis, $\unit{z}$, at a constant and prescribed (i.e., known) rate $\Omega_\mathrm{out}$ and an angle $\alpha$. Our conclusions are as follows.

\medskip \noindent 1. Extremely high eccentricities, $e_\mathrm{max}\rightarrow 1$, can be attained in the inner orbit (i.e., the inner planetary orbit), even if the initial inclination of the inner orbit with respect to the outer orbit is small and less than the minimum LK angle (i.e., $<40^\circ$). These enhancements occur already at the quadrupole order. 

\medskip \noindent 2. The nature of the eccentricity enhancement depends sensitively on the (dimensionless) quantity $\beta \equiv \Omega_\mathrm{out} t_\mathrm{LK}$, where $t_\mathrm{LK}$ is the LK time-scale of the inner orbit driven by the outer orbit.

The case $\beta=0$ corresponds to the unperturbed three-body problem, for which the maximum eccentricity is given by the canonical LK expression, equation~(\ref{eq:e_max_beta_zero}). High eccentricities can be attained, but only if $i_0$ is high.

In the limit of $\beta \gg 1$, we obtained an analytic result for $e_\mathrm{max}$ by averaging over the precessional cycle of the outer orbit (equation~\ref{eq:e_max_inf_beta}). Effectively, the classical LK result for the maximum eccentricity can be used, replacing the initial inclination between the inner and outer orbits by the initial inclination of the inner orbit with respect to the $\unit{z}$ axis, i.e., the axis around which $\unit{L}_\mathrm{out}$ is precessing.

Most importantly, if $\beta\sim 1$, then large $e_\mathrm{max}$ can be achieved for modest initial inclinations. In particular, $e_\mathrm{max}\rightarrow 1$ for $i_0 \gtrsim 50^\circ$ if $\alpha=5^\circ$ (see \F\,\ref{fig:fig3_emax_i_rel}). For $\alpha=5^\circ$, there is a complicated dependence of $e_\mathrm{max}$ on $i_0$, with chaotic `ridges' occurring at specific $i_0$ and $0.26\lesssim \beta \lesssim 0.7$, and a broad envelope of maximum eccentricities appearing for $0.7\lesssim \beta\lesssim 3$. For larger $\alpha$, there are fewer ridges, and a large chaotic region appears for $\beta$ around 1 (see \F\,\ref{fig:fig3_emax_i_rel_30}, in which $\alpha=30^\circ$).

\medskip \noindent 3. The eccentricity excitations around $\beta \sim 1$ can be explained by the enhanced maximum inclination $i_\mathrm{max}$ as a result of the precessing $\unit{L}_\mathrm{out}$, which subsequently drives strong LK evolution. This was illustrated in \F\,\ref{fig:fig3_emax_i_rel}, in which the maximum eccentricities were computed numerically from the maximum inclinations using the ad hoc expression equation~(\ref{eq:e_max_i_max_ad_hoc}).

\medskip \noindent 4. We briefly considered how the above is modified with the addition of short-range forces, in particular, general relativistic precession (\S\,\ref{sect:num_gen:SRF}). Although the maximum eccentricities are limited by general relativistic precession, there is still significant enhancement of the eccentricity compared to $\beta=0$ for small inclinations. 

\medskip \noindent 5. To further explore point (3), we considered a simplified model, the `inclination model' (\S\,\ref{sect:incl_model}), in which we set $e=0$ for the inner orbit. This simple model can be studied semi-analytically, and correctly predicts that high mutual inclinations are attained even if $i_0$ is small, provided that $\beta$ is around unity. For example, even if $i_0$ is as small as $2^\circ$, $i_\mathrm{max}$ can reach values of $\approx 70^\circ$ if $\beta$ is near 0.5 and $\alpha=5^\circ$. For a large range of inclinations, a good estimate for the minimum value of $\beta$ for which high mutual inclinations can be achieved is given by equation~(\ref{eq:beta_crit}). Typically, the inclination enhancement drops off rapidly as $\beta$ increases beyond $\beta_\mathrm{crit}$, with little enhancement occurring for $\beta \gtrsim 2$ (see \F\,\ref{fig:i_e_max_beta}). The  behaviour of the eccentricity excitation around $\beta\sim 1$ for the general model can be crudely reproduced if the analytic ad hoc LK expression is used to compute $e_\mathrm{max}$ from the maximum inclination obtained in the semi-analytic inclination model.

\section*{Acknowledgements}
We thank the referee, Cristobal Petrovich, for insightful and helpful comments. We also thank Scott Tremaine for comments on the manuscript, and Fabio Antonini, Evgeni Grishin and Ben Bar-Or for simulating discussions. ASH gratefully acknowledges support from the Institute for Advanced Study, and NASA grant NNX14AM24G. DL has been supported in part by NASA grants NNX14AG94G and NNX14AP31G, and a Simons Fellowship from the Simons Foundation.

\bibliographystyle{mnras}
\bibliography{literature}

\appendix
\section{Maximum eccentricity as a function of the initial inclination for $\alpha=30^\circ$}
\label{app:alpha_30}
\begin{figure*}
\centering
\includegraphics[scale = 0.42, trim = 35mm 0mm 0mm 0mm]{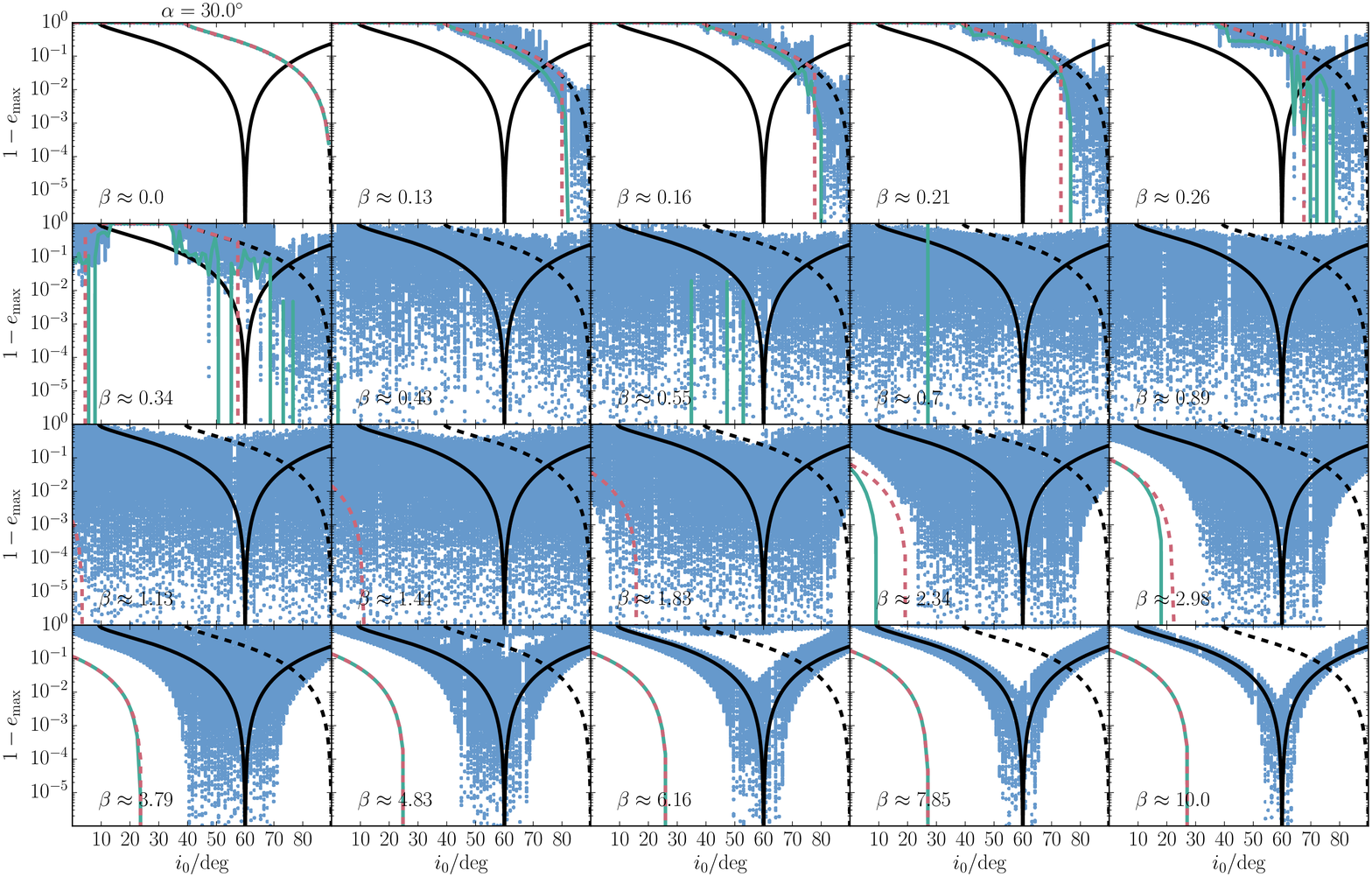}
\caption{\small Similar to \F\,\ref{fig:fig3_emax_i_rel}, now with $\alpha=30^\circ$. Blue points: the (local) maximum eccentricities as a function of initial inclination $i_0$. The black dashed lines show the canonical result (equation~\ref{eq:e_max_beta_zero}), which applies in the limit $\beta=0$. The solid black lines show equation~(\ref{eq:e_max_inf_beta}), which is valid in the limit $\beta\gg1$. The solid green lines show the maximum eccentricities computed from the numerically determined maximum inclinations and the ad hoc equation~(\ref{eq:e_max_i_max_ad_hoc}) (see \S\,\ref{sect:num_gen:intermediate}). Similarly, the red dashed lines show the maximum eccentricities computed using equation~(\ref{eq:e_max_i_max_ad_hoc}) with $i_\mathrm{max}$ determined from the inclination model (see \S\,\ref{sect:incl_model:maxi}). }
\label{fig:fig3_emax_i_rel_30}
\end{figure*}

In \F\,\ref{fig:fig3_emax_i_rel_30}, we show a figure similar to \F\,\ref{fig:fig3_emax_i_rel}, now with $\alpha=30^\circ$. The region in the parameter space for which large eccentricities are reached is now larger: $e_\mathrm{max}\rightarrow1$ even for small $i_0$, in the range $0.43 \lesssim \beta \lesssim 1.44$. The inclination model predicts $i_\mathrm{max}\geq90^\circ$ in this regime, hence we set $e_\mathrm{max} = 1$ (see the discussion in \S\,\ref{sect:num_gen:intermediate}). Note that in the limit of large $\beta$, the inclination model correctly predicts $i_\mathrm{max}$ (compare the solid green and red dashed lines in \F\,\ref{fig:fig3_emax_i_rel_30}); however, in this limit, the maximum eccentricity should be computed from $i_{z,0}$ (as shown with the solid black lines in \F\,\ref{fig:fig3_emax_i_rel_30}; see \S\,\ref{sect:num_gen:large} and equation~\ref{eq:e_max_inf_beta}). 

\label{lastpage}
\end{document}